\documentclass[ALICE,manyauthors]{cernphprep}
\usepackage[english]{babel}
\usepackage{graphicx}
\usepackage{verbatim}
\usepackage{amsfonts}
\usepackage{makeidx}
\usepackage{color}
\usepackage{amsmath}
\usepackage{lineno}
\usepackage{epstopdf}
\usepackage{multirow}
\usepackage{hyperref}
\usepackage{cite}

\newcommand{\sqrts}{\sqrt{s}}
\newcommand{\sqrtsNN}{\sqrt{s_{\rm \scriptscriptstyle NN}}}
\newcommand{\lsim}{\,{\buildrel < \over {_\sim}}\,}
\newcommand{\gsim}{\,{\buildrel > \over {_\sim}}\,}
\newcommand{\av}[1]{\left\langle #1 \right\rangle}

\newcommand{\GeV}{\mathrm{GeV}}
\newcommand{\TeV}{\mathrm{TeV}}

\newcommand{\gev}{\mathrm{GeV}}
\newcommand{\tev}{\mathrm{TeV}}

\newcommand{\mm}{\mathrm{mm}}
\newcommand{\cm}{\mathrm{cm}}

\newcommand{\mum}{\mathrm{\mu m}}

\renewcommand{\d}{\mathrm{d}}

\newcommand{\mub}{\mathrm{\mu b}}

\newcommand{\PbPb}{\mbox{Pb--Pb}}
\newcommand{\Npart}{N_{\rm part}}
\newcommand{\Ncoll}{N_{\rm coll}}
\newcommand{\Raa}{R_{\rm AA}}
\newcommand{\RAA}{R_{\rm AA}}
\newcommand{\TAA}{T_{\rm AA}}

\newcommand{\pt}{p_{\rm t}}

\newcommand{\DtoKpi}{{\rm D^0\to K^-\pi^+}}
\newcommand{\DtoKpipi}{{\rm D^+\to K^-\pi^+\pi^+}}
\newcommand{\DstartoDpi}{{\rm D^{*+}\to D^0\pi^+}}
\newcommand{\Dzero}{{\rm D^0}}
\newcommand{\Dstar}{{\rm D^{*+}}}
\newcommand{\Dplus}{{\rm D^+}}

\newcommand{\dEdx}{{\rm d}E/{\rm d}x}
\newcommand{\mufac}{\mu_{\rm F}}
\newcommand{\muren}{\mu_{\rm R}}

\begin{document}

\begin{titlepage}

\PHnumber{2012-069}                 
\PHdate{12 October 2012}              

\title{Suppression of high transverse momentum $\rm D$ mesons\\ in central Pb--Pb collisions at  $\sqrtsNN=2.76~\rm TeV$}

\Collaboration{ALICE Collaboration%
         \thanks{See Appendix~\ref{app:collab} for the list of collaboration 
                      members}}
\ShortAuthor{ALICE Collaboration}
\ShortTitle{Suppression of high $\pt$ $\rm D$ mesons in Pb--Pb collisions at  $\sqrtsNN=2.76~\rm TeV$}

\begin{abstract}
The production of the prompt charm mesons $\Dzero$, $\Dplus$, $\Dstar$, 
and their antiparticles, 
was measured with the ALICE detector in Pb--Pb collisions at the LHC, at a 
centre-of-mass energy $\sqrtsNN=2.76~\tev$ per nucleon--nucleon collision. 
The $\pt$-differential production yields in the range $2<\pt<16$~$\gev/c$
at central rapidity, $|y|<0.5$, were used to calculate the nuclear 
modification factor $\RAA$ with respect to a proton--proton reference 
obtained from the cross section measured at $\sqrts=7~\tev$ and scaled 
to $\sqrts=2.76~\tev$. 
For the three meson species, $\RAA$ shows a suppression by a factor 3--4, 
for transverse momenta larger than $5~\gev/c$
in the 20\% most central collisions.
The suppression is reduced for peripheral collisions.
\end{abstract}

\end{titlepage}

 \setcounter{page}{2}

\section{Introduction}
\label{sec:intro}
A high-density colour-deconfined state of strongly-interacting matter is expected to be formed
in high-energy collisions of heavy nuclei. 
According to calculations of Quantum Chromodynamics (QCD) on the lattice, 
under the conditions of high energy density and 
temperature reached in these collisions, a phase transition to a Quark-Gluon Plasma (QGP) occurs.
In such conditions, the confinement of quarks and gluons 
into hadrons vanishes, and chiral symmetry is restored 
(see e.g.~\cite{karsch,Borsanyi,Bazavov}).
Heavy-flavour hadrons, containing charm and beauty,
are effective probes of the conditions of the medium formed 
in nucleus--nucleus collisions at high energy. 
Hard partons, including gluons, light-flavour quarks, and heavy quarks, 
are produced at the initial stage of the collision in high-virtuality scattering
processes. They interact with the medium, and are expected to be sensitive to its energy density,
through the mechanism of parton energy loss. This QCD energy loss is expected to occur
via both inelastic (medium-induced gluon radiation, or radiative energy loss)~\cite{gyulassy,bdmps} and elastic (collisional energy loss)~\cite{thoma}
processes.
In QCD, quarks have a smaller colour coupling factor with respect to gluons, 
so that the energy loss for quarks is expected to be smaller than for gluons. 
In addition, the `dead-cone effect' should reduce small-angle gluon radiation 
for heavy quarks with moderate 
energy-over-mass values~\cite{dk,asw,dg,wang,whdg}, thus further attenuating 
the effect of the medium. 
Instead, other mechanisms, such as  
in-medium hadron formation and dissociation~\cite{adil,vitev}, would determine a
stronger effect on heavy-flavour hadrons, characterized by smaller formation 
times than light-flavour hadrons.
Finally, low-momentum heavy quarks may be to some extent thermalized in the 
hot and dense system through rescatterings and in-medium resonant 
interactions~\cite{rapp}.
 
One of the observables that are sensitive to the interaction of hard partons with the medium is the nuclear modification factor $R_{\rm AA}$. This quantity is defined as 
the ratio of particle production measured in nucleus--nucleus (AA)
to that expected from the proton--proton (pp) spectrum scaled by the 
average number $\av{N_{\rm coll}}$ of binary 
nucleon--nucleon collisions occurring in the nucleus--nucleus collision. 
Using the nuclear overlap function, which is defined as the convolution 
of the nuclear density profiles of the colliding ions in the Glauber 
model~\cite{glauber}, the nuclear modification factor 
of the transverse momentum ($\pt$) distribution can be expressed as: 
\begin{equation}
\label{eq:Raa}
R_{\rm AA}(\pt)=
{1\over \av{T_{\rm AA}}} \cdot 
{\d N_{\rm AA}/\d\pt \over 
\d\sigma_{\rm pp}/\d\pt}\,,
\end{equation}
where the AA spectrum corresponds to a given collision-centrality class and 
$\av{T_{\rm AA}}$ is the average nuclear overlap function for that
centrality class and is proportional to $\av{\Ncoll}$. 
In-medium energy loss determines a suppression, $\RAA<1$, of hadrons at 
moderate-to-high transverse momentum ($\pt\gsim 2~\gev/c$).
Given the aforementioned properties of parton energy loss, 
in the range $\pt\lsim 10~\gev/c$ where the heavy-quark masses are not negligible with respect to their momenta, 
an increase of
the $\RAA$ value (i.e.\ a smaller suppression) is expected when going from 
the mostly gluon-originated light-flavour hadrons (e.g.\ pions) 
to D and B mesons (see e.g.~\cite{whdg,adsw}): 
$\RAA^\pi<\RAA^{\rm D}<\RAA^{\rm B}$. 
The measurement and comparison of these different medium probes 
should provide a unique test of the 
colour-charge and mass dependence of parton energy loss.

Experiments at the Relativistic Heavy Ion Collider (RHIC) measured a strong 
suppression,
by a factor 4--5 at $\pt>5~\gev/c$, for light-flavour hadrons in central Au--Au 
collisions at $\sqrtsNN=200~\gev$~\cite{whitepapers}. 
An even stronger suppression ---up to a factor 7 at $\pt\approx 6$--$8~\gev/c$--- was observed in 
central Pb--Pb collisions at $\sqrtsNN=2.76~\tev$ at 
the Large Hadron Collider (LHC)~\cite{raapaper,chargedRAA,CMSraa}.
At RHIC, the suppression of heavy-flavour 
hadrons, measured indirectly from their inclusive decay 
electrons~\cite{phenixRAAe,starRAAe}, was found to be compatible with that of pions
and generally stronger than most expectations based on radiative energy loss~\cite{aswRHICe1,aswRHICe2}. 
At the LHC, a measurement by the CMS Collaboration indicates a 
strong suppression, by a factor about 3, in the nuclear modification factor of non-prompt J/$\psi$ particles
from B meson decays~\cite{CMSquarkonia}.

We present the first measurement of the nuclear modification factor for 
$\rm D^0$, $\rm D^+$, $\rm D^{*+}$ mesons, and their antiparticles, 
in Pb--Pb collisions at $\sqrtsNN=2.76~\tev$, 
carried out using the ALICE detector. 
The experimental apparatus~\cite{aliceJINST} 
is briefly presented in Section~\ref{sec:detector}, where
the Pb--Pb data sample used for this analysis is also described. The D meson signals are 
extracted using a selection based on displaced decay vertex reconstruction and particle 
identification of the decay products, as presented in Section~\ref{sec:signal}. The corrections 
applied to obtain the $\pt$-differential production yields, and the estimation of the systematic
uncertainties are described in Sections~\ref{sec:corrections} and~\ref{sec:systematics}, 
respectively.
The production of D mesons was measured in proton--proton collisions at 
$\sqrt{s}=7~\tev$ and compared to perturbative QCD (pQCD) predictions~\cite{Dpp7paper}. 
The reference for the $\RAA$ measurements was obtained 
by scaling these results to the Pb--Pb energy via a pQCD-driven approach
and was validated by comparing to data from a limited-statistics pp sample at this energy~\cite{Dpp2.76}. 
This is discussed in Section~\ref{sec:reference}. 
The results on the $\Dzero$, $\Dplus$, and $\Dstar$ 
nuclear modification factors as a function of transverse momentum 
and collision centrality are presented in Section~\ref{sec:results}. The results are compared
to the charged particle $\RAA$ measured with the ALICE detector~\cite{chargedRAA}, 
to the non-prompt J/$\psi$ results by the CMS Collaboration~\cite{CMSquarkonia},
and to model predictions.


\section{Experimental apparatus, data sample, event reconstruction and selection}
\label{sec:detector}
The ALICE detector, described in detail in~\cite{aliceJINST}, consists of 
a central barrel composed of various detectors for particle reconstruction 
at midrapidity, a forward muon spectrometer, 
and a set of forward detectors for triggering and event characterization.
In the following, the subsystems that are utilized in the D meson 
analysis will be briefly described.
In particular, the Inner Tracking System (ITS), the Time Projection 
Chamber (TPC), and the Time Of Flight (TOF) detector provide charged 
particle reconstruction and identification in the central pseudo-rapidity 
region ($|\eta|<0.9$).
They are embedded in a 0.5~T magnetic field parallel to the LHC beam 
direction ($z$-axis in the ALICE reference frame).
The VZERO detector and the Zero Degree Calorimeters (ZDC) are 
used for triggering and event selection, and the T0 detector to measure the 
start time (event time-zero) of the collision.

The data from $\PbPb$ collisions at centre-of-mass energy $\sqrtsNN=2.76~\tev$
used for this analysis were recorded in 
November and December 2010 during the first run with heavy-ions at the LHC.
The events were collected with an interaction trigger based on the information 
of the Silicon Pixel Detector (SPD) and the VZERO detector.
The SPD is the innermost part of the ITS.
It consists of two cylindrical layers of silicon pixel detectors located
at radial positions of 3.9 and 7.6~cm from the beam line, covering
the pseudo-rapidity ranges $|\eta|< 2.0$ and  $|\eta|< 1.4$, respectively.
The SPD contributes to the minimum-bias trigger if hits are detected on at 
least two different chips (each covering a detector area of 
$1.28\times1.41~\cm^2$) on the outer layer.
The VZERO detector is composed of two arrays of scintillator tiles covering
the full azimuth in the pseudo-rapidity regions $2.8 < \eta < 5.1$ (VZERO-A)
and $-3.7 < \eta < -1.7$ (VZERO-C). 
The events used in this analysis were collected with
two different interaction trigger configurations: in the first part of the 
data taking period, signals in two out of the three triggering detectors 
(SPD, VZERO-A, VZERO-C) were required, while in the second part a coincidence 
between the VZERO-A and VZERO-C detectors was used.
Events were further selected offline to remove background coming from parasitic
beam interactions on the basis of the timing information provided by the 
VZERO and the neutron ZDC detectors (two calorimeters located at 
$z\approx\pm 114$~m from the interaction point).
It was verified that the timing information from the ZDCs was available in all 
the hadronic interactions that passed the trigger condition.
The luminous region had an r.m.s.\ width of about 6~cm in the longitudinal 
direction and 50--$60~\mum$  in the transverse direction. These values
were stable during the entire data taking period.
Only events with a vertex found within $\pm$10~cm from the centre
of the detector along the beam line were considered for the D meson signal
extraction.

Collisions were classified according to their centrality, defined in terms of 
percentiles of the hadronic Pb--Pb cross section and determined from the
distribution of the summed amplitudes in the VZERO scintillator tiles.
To obtain the total hadronic cross section, this distribution was fitted 
using the Glauber model for the geometrical 
description of the nuclear collision~\cite{glauber} complemented
by a two-component model for particle 
production~\cite{ALICE-PbPbMult1,ALICE-PbPbMult2}.
The fit was performed in a range of measured VZERO amplitudes where the 
trigger is fully efficient for hadronic interactions and the contamination
by electromagnetic processes is negligible~\cite{ALICE-PbPbMult2}. 
This range corresponds to 90$\pm$1\% of the total hadronic cross section.
The nuclear modification factor $\RAA$ was measured for 
$\Dzero$, $\Dplus$, and $\Dstar$ mesons as a function of transverse momentum for the centrality 
classes 0--20\% and 40--80\%. 
In order to study in more detail its centrality dependence,
$\Raa$ was also evaluated, for wide $\pt$ intervals, in narrower centrality 
classes: 0--10\%, 10--20\%, 20--40\%, 40--60\%, and 60--80\%. 
Table~\ref{tab:centbins} shows the average values of the number of 
participating nucleons $\langle\Npart\rangle$, and of the nuclear 
overlap function $\langle\TAA\rangle$ in these centrality classes.
In the centrality range considered in this analysis, 0--80\%, and for
both the configurations of the interaction trigger described above, 
the trigger and event selection are fully efficient for hadronic interactions, 
and the contamination by electromagnetic processes is negligible.

In total, $13\times 10^6$ $\PbPb$ collisions with centrality in the range 
0--80\% passed the selection criteria described above and were used in the 
analysis.
The corresponding integrated luminosity is 
$L_{\rm int}=2.12\pm0.07~\mub^{-1}$. 

\begin{table}[!t]
\centering
\caption{Average values of the number of participating nucleons, and of 
the nuclear overlap function
for the considered centrality classes, expressed as percentiles of the 
hadronic cross section.
The values were obtained with a Monte Carlo implementation of the Glauber 
model assuming an 
inelastic nucleon--nucleon cross section of 64 mb~\cite{ALICE-PbPbMult2}.}
\begin{tabular}{|c|c|c|c|} 
\hline 
\rule[-0.3cm]{0cm}{0.8cm} Centrality class & $\langle\Npart\rangle$ &  $\langle\TAA\rangle~(\rm mb^{-1})$\\
\hline
\phantom{0}0--20\%  & $308 \pm 3$  & $18.93 \pm 0.74$\\
40--80\% & $\phantom{0}46 \pm 2$   &  $\phantom{0}1.20 \pm 0.07$\\
\hline
\phantom{0}0--10\%  & $357 \pm 4$ & $23.48 \pm 0.97$\\
10--20\% & $261 \pm 4$  & $14.43 \pm 0.57$\\
20--40\% & $157 \pm 3$ &  $\phantom{0}6.85 \pm 0.28$\\
40--60\%  & $\phantom{0}69 \pm 2$  &  $\phantom{0}2.00 \pm 0.11$\\
60--80\% & $\phantom{0}23 \pm 1$ &  $\phantom{0}0.42 \pm 0.03$\\
\hline
\end{tabular}
\label{tab:centbins}
\end{table}

The trajectories of the D meson decay particles were reconstructed from 
their hits in the TPC and in the ITS.
The TPC~\cite{TPCpaper} provides track reconstruction
with up to 159 three-dimensional space points per track 
in a cylindrical active volume that covers the region $85<r<247~\cm$ 
and $-250<z<+250~\cm$ in the radial and longitudinal directions, respectively.
The ITS~\cite{ITSalign} consists of six cylindrical layers of silicon 
detectors with radii in the range between 3.9~cm and 43.0~cm.
Around the two innermost layers equipped with pixel detectors 
(SPD, described above), Silicon Drift Detectors (SDD) are used in the two
intermediate layers, while the two outermost layers are made of double-sided
Silicon Strip Detectors (SSD).
The alignment of the ITS sensor modules, which is crucial to achieve the 
high space point resolution needed in heavy flavour analysis, was performed
using survey information, cosmic-ray tracks, and pp data, with the methods 
described in~\cite{ITSalign}.
For the residual misalignment along the $r\phi$ coordinate, an r.m.s.\ of about $8~\mum$ 
for SPD and $15~\mum$ for SSD modules was estimated~\cite{ITSalign,RossiVertex}. 
For SDD, with the calibration level reached on the 2010 data sample, the space point resolution along $r\phi$ is $\approx 60~\mum$ 
for those modules that do not suffer from significant drift field non-uniformities. 
The residual misalignment is included in an effective way in the detector simulation by randomly displacing 
the ITS modules with respect to their ideal positions according to the estimated precision of the alignment.

\begin{figure}[!t]
\begin{center}
\includegraphics[width=0.48\textwidth]{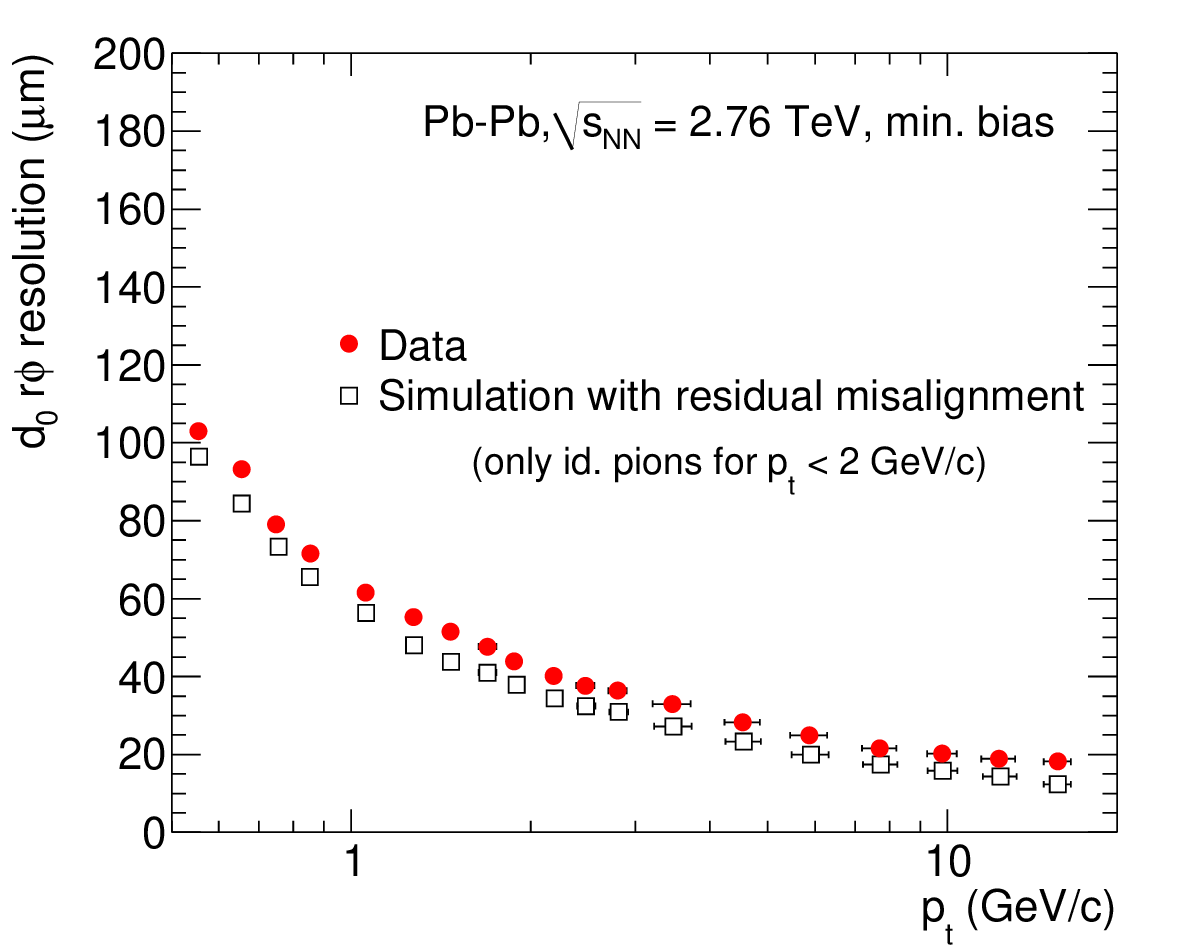}
\includegraphics[width=0.48\textwidth]{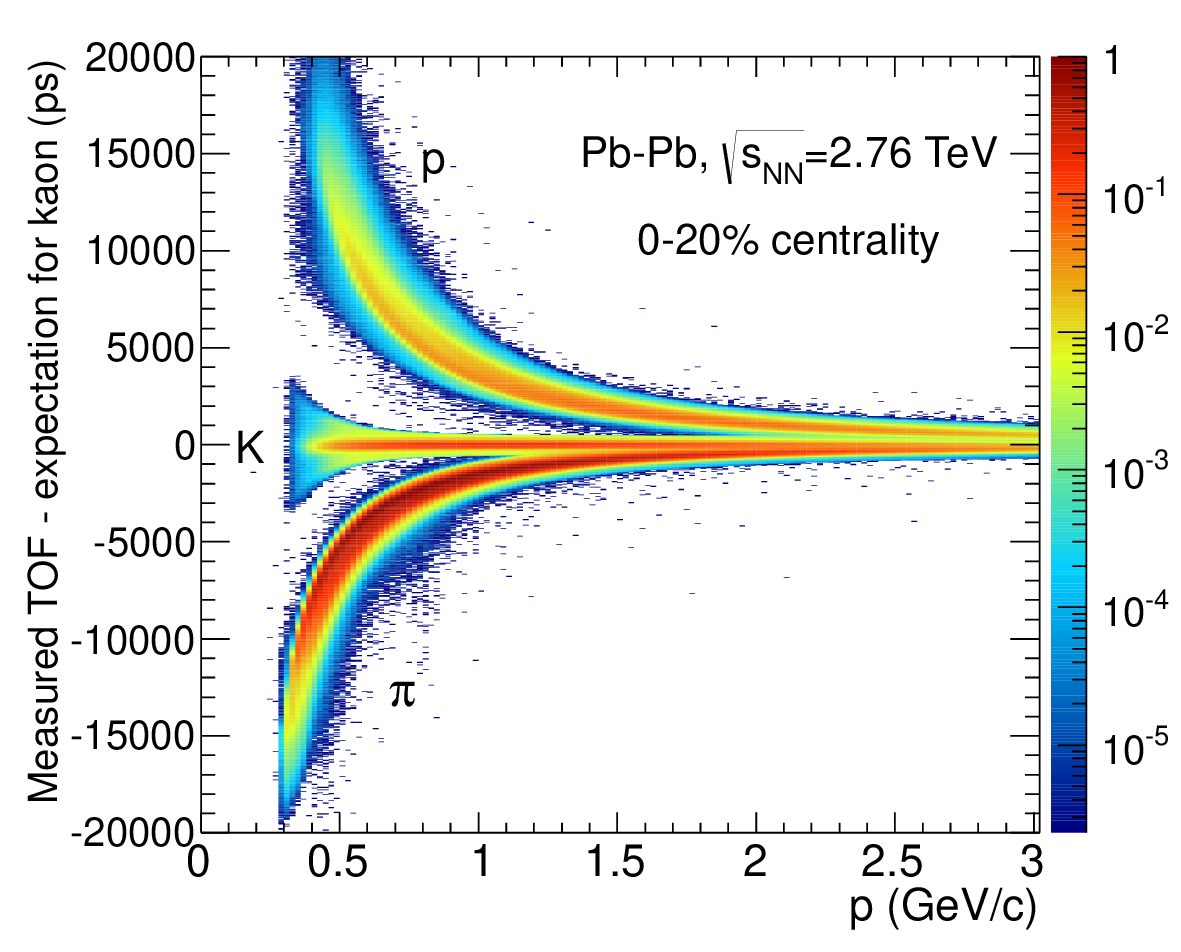}
\caption{Left: track impact parameter resolution in the transverse plane as 
a function of $\pt$ in Pb--Pb collisions. For $\pt<2~\gev/c$, pion 
identification by the TPC or TOF detectors is required; the results for data 
and simulation are shown. 
The simulation includes the effect of the residual geometrical misalignment of 
the sensor modules of the Inner Tracking System (see text for details). 
Right (colour online): difference between the measured time-of-flight and that 
expected under the kaon hypothesis as a function of track momentum for the 
20\% most central Pb--Pb collisions.}
\label{fig:imppar}
\end{center}
\end{figure}

The primary vertex position and covariance matrix were determined from the 
tracks reconstructed in the TPC and ITS 
by using an analytic $\chi^2$ minimization method, applied after 
approximating the tracks to straight lines in the vicinity of their common 
origin.
The same algorithm was used for the reconstruction of the decay
vertices of $\Dzero$ and $\Dplus$ candidates.
The high spatial resolution of the reconstructed hits, together 
with the low material budget (on average 7.7\% of a radiation length for 
the ITS at $\eta=0$) 
and the small distance of the innermost layer 
from the beam vacuum tube, allows for the measurement of the track impact 
parameter in the
transverse plane ($d_0$), i.e. the distance of closest approach of the 
track to the 
primary vertex along $r\phi$, with a resolution better than 65~$\mu$m for 
transverse momenta $\pt> 1$~GeV/$c$.
The impact parameter resolution $\sigma_{d_0}$ is shown in 
the left-hand panel of Fig.~\ref{fig:imppar} as a function 
of $\pt$, for data and simulation, for charged hadron tracks selected 
with the same criteria used in the D meson analysis.
The applied track quality cuts were based on the request of having
at least 70 associated space points (out of a maximum of 159) in the TPC with
a $\chi^2$ per degree-of-freedom of the momentum fit lower than 2, and 
at least 2 associated hits in the ITS, out of which at least one has to be
in the silicon pixel layers.
Only tracks with transverse momentum $\pt>0.5~\gev/c$ (0.7 for the 20\% most 
central collisions) and $|\eta|<0.8$ were used for the D meson analysis and 
are displayed in Fig.~\ref{fig:imppar} (left).
For $\pt<2~\gev/c$, only particles identified as pions were selected, 
as explained in the next paragraph.
The impact parameter resolution is better than for pp 
collisions~\cite{Dpp7paper}, e.g. by $\approx 10~\mum$ at $\pt=1~\gev/c$, 
since, in the Pb--Pb case, the primary vertex is reconstructed using a larger 
number of tracks, hence with better precision.
Indeed, the resolution on the transverse coordinates of the primary vertex 
is about 5~$\mum$ in central Pb--Pb collisions, while it is up to 40~$\mum$ 
in pp~\cite{Dpp7paper}.
The systematic effect on the D meson analysis of the small difference in 
resolution
($5~\mum$) between data and simulation
will be discussed in Section~\ref{sec:systematics}.  
The resolution on the transverse momentum of tracks reconstructed in the 
TPC and in the ITS, and passing the quality selection criteria described above, 
was measured to be about 1\% at $\pt=1~\GeV/c$ and about 2\% 
at $\pt=10~\GeV/c$.

The particle identification (PID) capabilities are provided by the measurement 
of the specific energy loss $\dEdx$ in the TPC and of the
time-of-flight in the TOF detector.
The $\dEdx$ samples measured by the TPC are reduced, by means of a truncated
mean, to a Gaussian distribution with a resolution of 
$\sigma_{\dEdx}/(\dEdx)\approx6\%$ which is slightly dependent on track 
quality and detector occupancy.
The TOF detector~\cite{TOFcommiss} is positioned at 370--399~cm from the beam 
axis and covers the full azimuth and the pseudo-rapidity range $|\eta|<0.9$.
In Pb--Pb collisions, in the centrality range 0--70\%, 
the overall time-of-flight resolution was measured to
be about 90~ps for pions with a momentum of 1~GeV/$c$.
This value includes the detector intrinsic resolution,
the electronics and calibration contribution,  the uncertainty on the 
start time of the event, and the tracking and momentum resolution. 
The start time of the event is measured by the T0 
detector, made of two arrays of Cherenkov counters located on either side of the 
interaction point and covering the pseudorapidity ranges 
$-3.28<\eta<-2.97$ and $4.61<\eta<4.92$, respectively.
For the events in which the T0 signal is not present, the start time is 
estimated using the particle arrival times at the TOF.
In the centrality class 70--80\%, the TOF resolution slightly worsens due to 
the increasing uncertainty on the start time determination, while still 
remaining below 100~ps. 

In the right-hand panel of Fig.~\ref{fig:imppar}, the difference 
between the measured time-of-flight and that expected under the kaon 
hypothesis is shown as a function of the track momentum for Pb--Pb collisions 
in the centrality range 0--20\%, illustrating the separation between pions, 
kaons and protons. 
The expected time-of-flight is calculated for the given mass hypothesis 
from the total integrated path length and the measured momentum of the track. 
A compatibility cut with the PID response from the TPC was used in 
order to decrease the contamination from tracks with wrong hit association 
in the TOF detector.
The bands corresponding to particles identified as pions (upper band), 
kaons (middle) and protons (lower band) are separated up to $p\approx2~\GeV/c$, 
corresponding to the momentum range in which the TOF PID is used in this 
analysis.

\section{D meson reconstruction and selection}
\label{sec:signal}
The $\Dzero$, $\Dplus$, and $\Dstar$ mesons and their antiparticles were 
reconstructed in the central rapidity region from their charged 
hadronic decay channels $\DtoKpi$ (with branching ratio, BR, 
of $3.87\pm 0.05\%$ and mean proper decay length $c\tau\approx 123~\mum$), 
$\DtoKpipi$ (BR of $9.13\pm 0.19\%$, 
$c\tau\approx 312~\mum$), and $\DstartoDpi$ 
(strong decay with BR of $67.7\pm 0.5\%$)~\cite{pdg}.
The D meson yields were extracted from an invariant mass analysis
of fully reconstructed decay topologies displaced with respect to the primary 
vertex, using the same procedure as for pp collisions~\cite{Dpp7paper}.

$\Dzero$ and $\Dplus$ candidates were defined from pairs and triplets of tracks 
with 
proper charge sign combination and selected by requiring
at least 70 associated space points in the TPC, with $\chi^2/{\rm ndf}<2$,
and at least 2 associated hits in the ITS, out of which at least one in the 
SPD.
A fiducial acceptance cut $|\eta|<0.8$ was applied as well, along with a 
transverse momentum threshold $\pt>0.5~\gev/c$ (0.7 for the 20\% most central 
collisions), aimed at reducing the large combinatorial background. 

$\Dstar$ candidates were obtained by combining the $\Dzero$ candidates
with tracks selected with transverse momentum 
$\pt>0.2~\gev/c$ in the centrality range 0--20\% and 
$\pt>0.1~\gev/c$ in 20--80\%.
The momentum of the pion from the $\Dstar$ decay is typically low, because of 
the small mass difference between the $\Dstar$ and $\Dzero$ mesons.
In order to reduce the combinatorics, these tracks were selected 
requiring at least 3 associated 
ITS hits (4 in the 0--20\% centrality class), in addition to the same TPC 
quality selection as that used for the $\Dzero$ and $\Dplus$ decay tracks.
In the centrality class 40--80\%, also tracks reconstructed only in the ITS, 
with at least 3 hits, were used 
to enhance the $\Dstar$ signal at low $\pt$.

The selection of the $\Dzero$ and $\Dplus$ decays 
was based on the reconstruction of secondary vertex topologies, with a 
separation of a few hundred microns from the interaction point. 
In the case of the $\Dstar$ decay, the secondary vertex topology 
of the produced $\Dzero$ was reconstructed.  
The selection is essentially the same as that used for the pp 
case~\cite{Dpp7paper} and exploits the separation between the secondary and 
primary vertices (decay length) and the pointing of the reconstructed 
meson momentum to the primary vertex.
The pointing condition is applied by requiring a small value for the 
angle $\theta_{\rm pointing}$ between the directions of
the reconstructed momentum of the candidate and of its flight line, defined 
by the positions of the primary and secondary vertices.
In order to cope with the much larger combinatorial background and to exploit 
the better resolution on the reconstructed primary vertex position, the cuts 
were in general tightened with respect to the pp case.
Two additional cuts, on the projections of the 
pointing angle and of the decay length in the transverse plane
($\theta_{\rm pointing}^{xy}$ and $L^{xy}$), were 
introduced to further suppress the combinatorial background. 

The cuts were defined so as to have large statistical 
significance of the signal and to keep the selection efficiency as high as 
possible.
This latter requirement was dictated also by the fact that too tight cuts 
result in an increased contribution to the raw yield from feed-down D mesons 
originating from decays of B mesons.
It was also checked that background fluctuations were not causing
a distortion in the signal line shape by verifying that the D meson mass and 
its resolution were in agreement with the PDG value and the simulation 
results, respectively.
The resulting cut values depend on the D meson $\pt$ and on the
centrality of the event.
They lead to a selection efficiency that increases with increasing $\pt$ and
decreases from peripheral to central collisions: looser cuts could be used for 
peripheral events, where the combinatorial background is lower.
The cut values quoted in the following refer to the tightest selections  
in the lower $\pt$ intervals for the 0--20\% centrality class.

The PID selection relies on the pion and kaon identification by the TPC and 
TOF detectors. Cuts at $\pm3\,\sigma$ 
around the expected mean energy deposit $\dEdx$ and time-of-flight were used.
This selection provides a strong reduction, by a factor of about 3, 
of the combinatorial background in the low-$\pt$ region, while
preserving most of the signal ($\approx 95\%$ according to simulations, 
as detailed in the next Section). 
In the $\Dstar$ case, a tighter PID cut at $2\,\sigma$ on the TPC $\dEdx$ was 
applied to the $\Dzero$ decay products in the centrality class 0--20$\%$, 
in order to cope with the large combinatorial background.

With the track selection described above, the acceptance in 
rapidity for D mesons drops steeply to zero for $|y|\gsim 0.5$ at low $\pt$ 
and $|y|\gsim 0.8$ for $\pt\gsim 5~\gev/c$. 
A $\pt$ dependent fiducial acceptance cut was therefore applied on the D meson
rapidity, $|y|<y_{\rm fid}(\pt)$, with $y_{\rm fid}(\pt)$ increasing from 
0.5 to 0.8 in $0<\pt<5~\gev/c$ according to a second order polynomial function
and taking a constant value of 0.8 for $\pt>5~\gev/c$. 

For $\Dzero$ mesons, the two decay tracks were selected requiring a 
significance of the impact parameter with respect to the 
event primary vertex $|d_0|/\sigma_{d_0}>0.5$ 
and a maximum distance of closest approach between each other of $250~\mum$.
The minimum decay length was set at $100~\mum$.
Furthermore, the cuts $d_0^{\pi}\times d_0^{\rm K}< - 45000~\mum^2$  
on the product 
of the decay track impact parameters and $L^{xy}/\sigma_{L^{xy}}>7$ on the
significance of the projection of the decay length in the transverse plane
were applied. 
A selection on the angle $\theta^*$ between the kaon momentum in 
the $\Dzero$ rest frame and the boost direction was used to reduce the 
contamination of background candidates that do not represent real two-body 
decays and typically have large values of $|\cos\theta^*|$.
The applied cut was $|\cos\theta^*|<0.8$.
The pointing of the $\Dzero$ momentum to the primary vertex 
was imposed via the cuts $\cos\theta_{\rm pointing}> 0.95$ and
$\cos\theta_{\rm pointing}^{xy}> 0.998$.

For $\Dplus$ mesons, a decay length of at least $1.9~\mm$ was required.
It should be noted that $\Dplus$ mesons in the 0--20\% centrality class
are reconstructed only for $\pt>6~\gev/c$, where the Lorentz dilation of 
the $\Dplus$ lifetime allows for a tight cut on the decay length.
Further requirements to reduce the combinatorial background were 
$\cos\theta_{\rm pointing}>0.995$, $\cos\theta_{\rm pointing}^{xy}> 0.997$, 
$L^{xy}/\sigma_{L^{xy}}>12$, and $\sum d_0^2 > (300~\mum)^2$ 
(sum of the squared impact parameters of the three decay tracks). 
The $\Dplus$ cuts are in general tighter than the $\Dzero$ ones because of the
larger combinatorial background.

\begin{figure}[!t]
\begin{center}
\includegraphics[width=\textwidth]{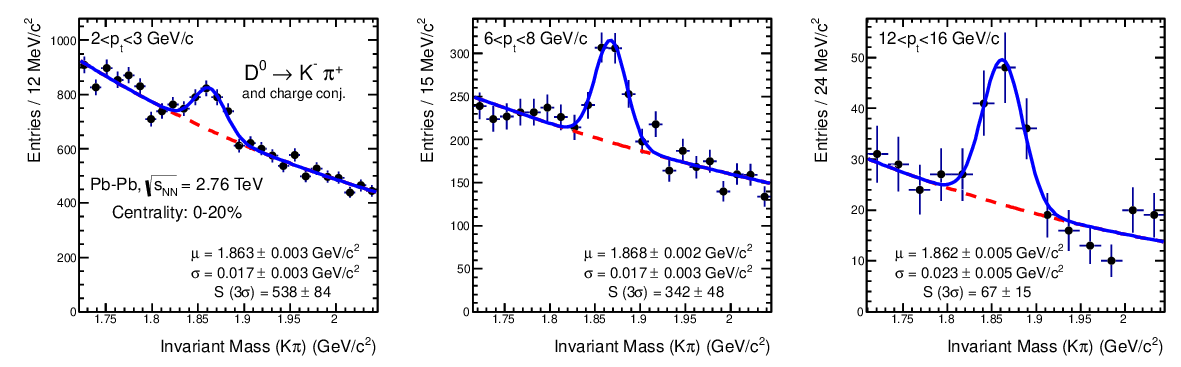}
\includegraphics[width=\textwidth]{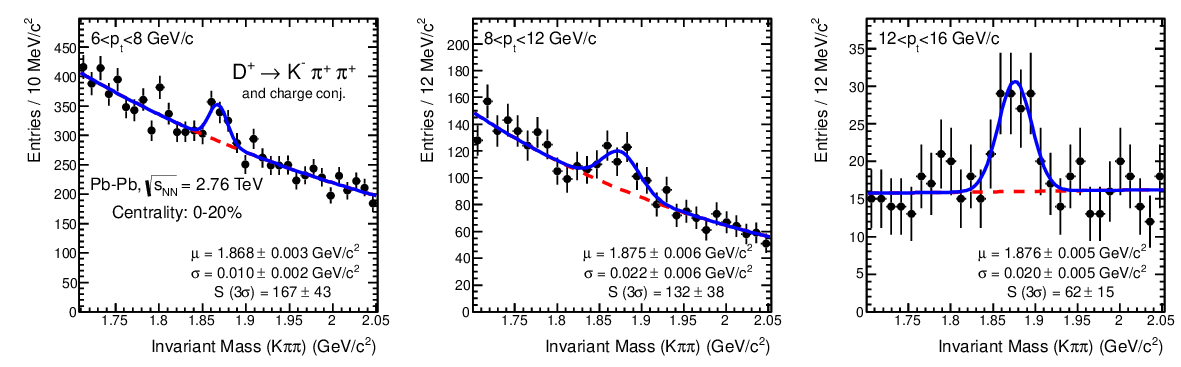}
\includegraphics[width=\textwidth]{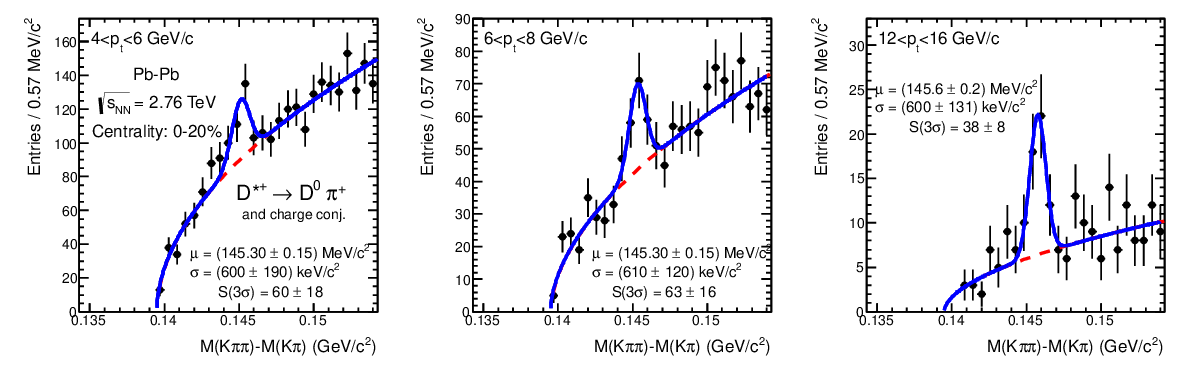} 
\caption{Invariant mass distributions for $\Dzero$ (upper panels), 
$\Dplus$ (central panels), and $\Dstar$ (lower panels) candidates and their
charge conjugates in selected $\pt$ intervals for $3.2\times 10^6$ 0--20\% 
central Pb--Pb collisions.
The curves show the fit functions described in the text.
The values of mean ($\mu$) and width ($\sigma$) of the signal peak are 
reported in the plots together with the raw signal yield.
The uncertainties on the signal yields reported in the figures are statistical 
only.}
\label{fig:invmass} 
\end{center}
\end{figure}

In the $\Dstar$ analysis, the selection of the decay $\Dzero$ was similar 
to that used for the $\Dzero$ analysis, with a tighter cut on the pointing 
angle, $\cos\theta_{\rm pointing}> 0.99$.
The decay pion was selected with the track quality cuts described above and 
requiring a minimum $\pt$ that varied in the range
0.1--$1~\gev/c$ depending on the $\Dstar$ momentum and event centrality. 
In the 0--20\% centrality class and for $\Dstar$ transverse momentum below 
$6~\gev/c$, a $3\,\sigma$ compatibility cut with respect to the pion 
expectation values was applied to the measured $\dEdx$ and time-of-flight.

Figure~\ref{fig:invmass} shows the invariant mass distributions of the 
selected $\Dzero$, $\Dplus$, and $\Dstar$ candidates in some of 
the $\pt$ intervals used in the analysis, for the 0--20\% centrality class. 
 The $\Dzero$ and $\Dplus$ yields were extracted by fitting the distributions with a function composed of a Gaussian
for the signal and an exponential term that describes the background shape.  The $\Dstar$ background was described with a threshold function 
multiplied by an exponential~\cite{Dpp7paper}. 
The centroids of the Gaussians were found to be compatible
with the PDG masses of the D mesons~\cite{pdg},
and their widths to be 
well reproduced in the simulation. 
The signal yields (sum of particle and antiparticle) are reported in 
Table~\ref{tab:yields} for the $\pt$ intervals considered 
in the analysis, for the centrality classes 0--20\% and 40--80\%.

\begin{table}[!ht]
\caption{Measured raw yields for 
$\Dzero$, $\Dplus$, and $\Dstar$ mesons and their antiparticles
in the transverse momentum intervals considered for the 0--20\% and 40--80\% 
centrality classes.
The systematic uncertainty estimation is described in 
Section~\ref{sec:systematics}. }
\centering
\small
\renewcommand{\arraystretch}{1.1}
\begin{tabular}{|c|c|c|c|c|c|c|}
\hline
$\pt$ & \multicolumn{6}{c|}{$N^{\rm raw}~\pm {\rm stat.} \pm {\rm syst.}$}\\
 interval &  \multicolumn{3}{c|}{0--20\% centrality} 
& \multicolumn{3}{c|}{40--80\% centrality}\\
 (GeV/$c$) &  $\Dzero+\overline{\rm D}^0$ & $\Dplus+\rm D^-$& $\Dstar+\rm D^{*-}$
&  $\Dzero+\overline{\rm D}^0$ & $\Dplus+\rm D^-$ & $\Dstar+\rm D^{*-}$ \\
\hline
2--3 & $538 \pm \phantom{0}84 \pm 43$ & -- &--
     & $231 \pm 31 \pm 12$   & -- & \multirow{2}{*}{$82 \pm 21 \pm 12$}\\
\cline{1-6}
3--4 & $774 \pm 108 \pm 46$ & --  & --
     & $241 \pm 32 \pm 12$ & $\phantom{0} 58 \pm 19 \pm \phantom{0}9$ & \\
\hline
4--5 & $583 \pm \phantom{0}79 \pm 35$ & --  & \multirow{2}{*}{$60 \pm 18 \pm 12$}
     & $176 \pm 20 \pm \phantom{0}9$ & \multirow{2}{*}{$114 \pm 22 \pm \phantom{0}6$} & $36 \pm \phantom{0}7 \pm \phantom{0}5$\\
\cline{1-3}
\cline{5-5}
\cline{7-7}
5--6 & $318 \pm \phantom{0}67 \pm 19$  & -- &
     & $\phantom{0}87\pm 13 \pm \phantom{0}4$ & &$29 \pm \phantom{0}9 \pm \phantom{0}3$\\
\hline
6--8 & $342 \pm \phantom{0}48 \pm 21$ & $167 \pm 43\pm 33$ &$63 \pm 16 \pm \phantom{0}6$
     & $113 \pm 14 \pm \phantom{0}6$ & $130 \pm 34 \pm 20$ &$47 \pm 13 \pm \phantom{0}5$ \\
\hline
\phantom{1}8--12 & $327\pm \phantom{0}41\pm 20$ & $132\pm 38\pm 20$ &$55 \pm 12 \pm \phantom{0}6$ 
      & $107 \pm 15 \pm \phantom{0}6$ & $119 \pm 26 \pm 18$ &$57 \pm 11 \pm \phantom{0}6$ \\
\hline
12--16 & $\phantom{0}67 \pm \phantom{0}15\pm \phantom{0}7$ & $\phantom{0}62\pm 15\pm \phantom{0}6$ &$38 \pm \phantom{0}8 \pm \phantom{0}4$
       & $\phantom{0}41\pm \phantom{0}9 \pm \phantom{0}2$ & -- &$23 \pm \phantom{0}6 \pm \phantom{0}2$\\
%
\hline
\end{tabular}
\label{tab:yields}
\end{table}

\section{Corrections}
\label{sec:corrections}
The D meson raw yields extracted from the fits to the invariant mass distributions
were corrected to obtain the production yields for primary 
(i.e. not coming from weak decays of B mesons) 
$\Dzero$, $\Dplus$, and $\Dstar$.
The contribution of secondary D mesons from B decays, which is of the order
of 15\% as explained in the following, was estimated using pQCD predictions 
for B production and Monte Carlo simulations.
The D mesons remaining after the subtraction of the B feed-down contribution
are those produced at the interaction vertex, and they will be referred 
to as `prompt' in the following.

The prompt D meson production yields were calculated starting from
the raw yields ($N^{\rm raw}$, reported in the previous 
section) divided by a factor of two to evaluate the charge (particle and 
antiparticle) averaged yields.
These were corrected for the B meson decay 
feed-down contribution (i.e. multiplied by the prompt fraction 
$f_{\rm prompt}$), and divided by the acceptance-times-efficiency for prompt 
D mesons, $({\rm Acc}\times\epsilon)_{\rm prompt}$. 
They were normalized according to the decay channel 
branching ratio ({\rm BR}), $\pt$ interval width ($\Delta \pt$), rapidity coverage 
($\Delta y$), and the number of events analyzed ($N_{\rm evt}$). 
As an illustration, the ${\rm D^+}$ yields were computed as:
\begin{equation}
  \label{eq:dNdpt}
  \left.\frac{{\rm d} N^{\rm D^+}}{{\rm d}\pt}\right|_{|y|<0.5}=
  \frac{1}{ \Delta y \,\Delta \pt}\frac{\left.f_{\rm prompt}(\pt)\cdot \frac{1}{2} N^{\rm D^\pm~raw}(\pt)\right|_{|y|<y_{\rm fid}}}{({\rm Acc}\times\epsilon)_{\rm prompt}(\pt) \cdot{\rm BR} \cdot N_{\rm evt}}\,.
\end{equation}
As mentioned in Section~\ref{sec:signal}, the D meson yields were measured in 
a rapidity range varying 
from $|y|<0.5$ at low $\pt$ to $|y|<0.8$ at high $\pt$.
The rapidity acceptance correction factor $\Delta y=2\,y_{\rm fid}$ assumes
a uniform rapidity  distribution for D mesons in the measured $y$ range.
This assumption was checked to the $1\%$ level~\cite{Dpp7paper} with 
PYTHIA~\cite{Pythia} pp simulations with the Perugia-0 
tuning~\cite{Perugia0}.

The acceptance-times-efficiency corrections ${\rm Acc}\times\epsilon$ 
were obtained using Monte Carlo simulations. 
Minimum-bias Pb--Pb collisions at $\sqrtsNN=2.76~\tev$
were produced with the HIJING~v1.36 event generator~\cite{Hijing}.
Prompt and feed-down (B decays) D meson signals were added using pp events from the PYTHIA~v6.4.21 
event generator~\cite{Pythia} with the Perugia-0 tuning~\cite{Perugia0}. 
Each injected pp event
was required to contain a $\rm c\overline c$ or $\rm b\overline b$ pair and D mesons
were forced to decay in the hadronic channels of interest for the analysis. The number of 
pp events added to each Pb--Pb event was adjusted according to the Pb--Pb collision
centrality.
The simulations used the GEANT3~\cite{Geant} particle transport package together with a detailed description of the geometry of the apparatus and of the detector response.
The simulation was configured to reproduce the conditions of the luminous region
 and of all the ALICE subsystems, 
in terms of active electronic channels, calibration level, and their time evolution within the Pb--Pb data taking period. 

\begin{figure}[t!]
\begin{center}
\includegraphics[width=\textwidth]{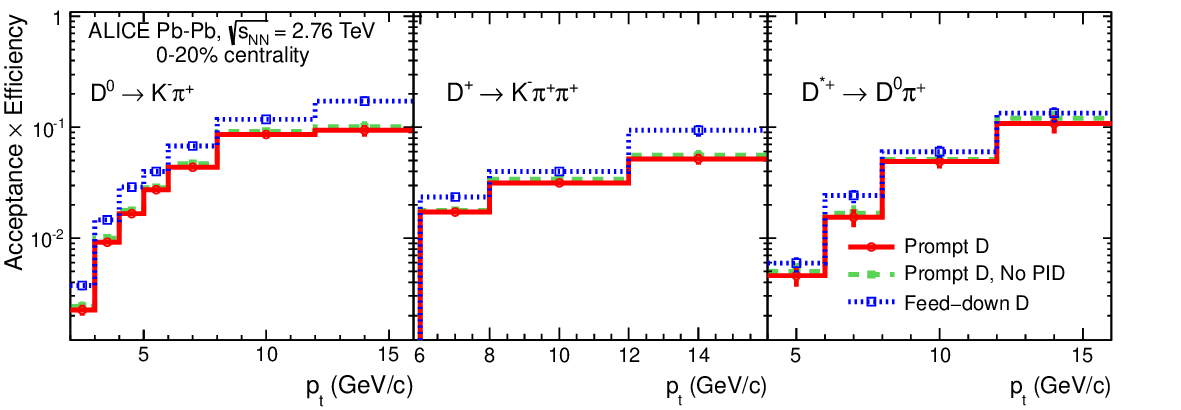}
\end{center}
\caption{Acceptance-times-efficiency in Pb--Pb collisions (0--20\% centrality class) 
for $\Dzero$ (left), $\Dplus$ (middle), and $\Dstar$ (right) mesons. 
The efficiencies for prompt (solid lines) and feed-down (dotted lines)
D mesons are shown. Also displayed, for comparison, the efficiency for 
prompt D mesons without PID selection (dashed lines).}
\label{figure:Eff_D0_Dplus}
\end{figure}

The efficiencies were evaluated in centrality classes corresponding
to those used in the analysis of the data in terms of charged-particle 
multiplicity, hence of detector occupancy.
Figure~\ref{figure:Eff_D0_Dplus} shows the $\DtoKpi$, $\DtoKpipi$, and 
$\DstartoDpi$ acceptance-times-efficiency for prompt and feed-down 
D mesons with rapidity $|y|<y_{\rm fid}$. 
The efficiencies correspond to Pb--Pb collisions in the 
centrality class 0--20\%.
The selection cuts described in Section~\ref{sec:signal} were 
applied. The values for the case of not applying PID are shown as well,
in order to point out that this selection is about 95\% efficient for the signal.
For the three meson species, the acceptance-times-efficiency increases 
with $\pt$, starting from few per mil and reaching $\approx 5$--10\% at high $\pt$.
No significant difference in the acceptance-times-efficiency for particles
and antiparticles was observed.

The acceptance-times-efficiencies for D mesons from B decays are larger than 
for prompt D mesons by a factor of approximately 2,
because the decay vertices of the feed-down D mesons 
are more displaced from the primary vertex and, thus, more efficiently
selected by the cuts. 

In the 40--80\% centrality class, as discussed in the
previous Section, the selection cuts were looser, resulting in a 
higher efficiency.
The dependence of the D meson selection efficiency on the detector occupancy 
was evaluated by comparing the efficiencies for central 
(0--20\% centrality class) and peripheral (40--80\% centrality class) events 
when applying the same selection cuts (those of the 0--20\% class were used 
as a reference). 
The results showed only small variations as a function of centrality, 
e.g. $\approx 5$--$10\%$ for $\Dzero$, as 
expected from the small variation of the single track reconstruction 
efficiency with centrality~\cite{raapaper}.
Indeed, also the efficiency of the topological selection is expected to 
be practically independent of centrality in the considered range 0-80\% where 
the resolution on the primary vertex position is not
significantly affected by the multiplicity of tracks used in its determination.

The prompt D meson production yields ${\rm d}N/{\rm d}\pt$ in Pb--Pb 
collisions were obtained by subtracting the contribution of D mesons 
from B decays with the same procedure used for the measurement of 
the production cross sections in pp collisions~\cite{Dpp7paper}.
In detail, the feed-down contribution was estimated using 
the beauty production cross section from the FONLL calculation~\cite{fonllpriv},
the B$\rightarrow$D decay kinematics from the EvtGen package~\cite{evtgen},
and the Monte Carlo efficiencies for feed-down D mesons. 
For Pb--Pb collisions, 
the FONLL feed-down cross section in pp at $\sqrt{s}=2.76~\tev$ 
was scaled by the average nuclear overlap function $\langle \TAA \rangle$ in each centrality class.
Thus, omitting for brevity the symbol of the $\pt$-dependence $(\pt)$, 
the fraction of prompt D mesons reads:
\begin{equation}
  \label{eq:fcNbMethod}
\begin{split}
f_{\rm prompt} &= 1-(N^{\rm D~feed-down~raw}/N^{\rm D~raw})=\\
 &= 1 - \langle \TAA \rangle 
	 \cdot \left( \frac{{\rm d}^2 \sigma}{{\rm d}y \, {\rm d}\pt } \right)^{{\sf FONLL}} _{{\rm feed-down}} \cdot \RAA^{\rm feed-down} \cdot \frac{({\rm Acc}\times\epsilon)_{\rm feed-down}\cdot\Delta y \, \Delta\pt \cdot {\rm BR} \cdot N_{\rm evt}  }{ N^{\rm D~raw }  / 2} \, ,
\end{split}
\end{equation}
where $({\rm Acc}\times\epsilon)_{{\rm feed-down}}$ is the 
acceptance-times-efficiency for feed-down D mesons.
The nuclear modification factor of the feed-down D mesons, $\RAA^{\rm feed-down}$,
is related to the nuclear modification of beauty production in Pb--Pb 
collisions, which is currently unknown. 
We therefore assumed for the correction that the nuclear modification factors 
for feed-down and prompt D mesons are equal 
($\RAA^{\rm feed-down}=\RAA^{\rm prompt}$) and varied this hypothesis
in the range $1/3<\RAA^{\rm feed-down}/\RAA^{\rm prompt}<3$ 
to determine the systematic uncertainty.
This hypothesis is justified by the range of the model predictions for the charm and beauty $\RAA$~\cite{whdg,adsw} and, as discussed in Section~\ref{sec:systematics}, by the CMS Collaboration results on $\RAA$ for 
non-prompt $\rm J/\psi$~\cite{CMSquarkonia}.
The value of $f_{\rm prompt}$ depends on the D meson species, the transverse
momentum interval, the applied cuts, the parameters used in the FONLL B 
prediction, and the hypothesis on $\RAA^{\rm feed-down}$.
The resulting values, for the case $\RAA^{\rm feed-down}=\RAA^{\rm prompt}$,
range from $\approx 0.95$ in the lowest transverse momentum interval 
($2<\pt<3~\gev/c$) to $\approx 0.85$ at high $\pt$.


\section{Reference pp cross section at $\sqrt{s}=2.76~\tev$}
\label{sec:reference}
The reference pp cross sections used for the determination of the nuclear 
modification factors were obtained by applying a $\sqrt{s}$-scaling~\cite{scaling} 
to the cross sections 
measured at $\sqrt{s}=7~\tev$~\cite{Dpp7paper}. 
The scaling factor for each D meson species
was defined as the ratio of the cross sections from 
the FONLL pQCD calculations~\cite{fonllpriv} at $2.76$ and $7~\tev$. 
The same values of the pQCD factorization scale $\mufac$ and renormalization 
scale $\muren$, 
and of the charm quark mass $m_{\rm c}$ were used in the calculation
for the different energies. Namely, $\mufac=\muren=m_{\rm t}$ with $m_{\rm t}=\sqrt{\pt^2+m_{\rm c}^2}$
and $m_{\rm c}=1.5~\gev/c^2$.
The theoretical uncertainty on the
scaling factor was evaluated by considering the envelope of the scaling factors resulting 
by varying independently the scales in the ranges  
$0.5<\mu_{\rm R}/m_{\rm t}<2$, $0.5<\mu_{\rm F}/m_{\rm t}<2$, 
with $0.5<\mu_{\rm R}/\mu_{\rm F}<2$, 
and the quark mass in the range $1.3<m_{\rm c}<1.7~\gev/c^2$, following the prescription in~\cite{FONLLrhic}. 
This uncertainty ranges from
$^{+30}_{-10}\%$ at $\pt=2~\gev/c$ to about $\pm 5\%$ for $\pt>10~\gev/c$~\cite{scaling}.
The procedure was validated by scaling the ALICE pp data 
to the Tevatron energy, $\sqrt{s}=1.96~\tev$, and comparing 
to CDF measurements~\cite{scaling,cdf}.
In addition, it was verified that the scaling factor and its uncertainty are the same if the GM-VFNS 
calculation~\cite{gmvfns} is used instead of FONLL~\cite{scaling}.

The $\rm D^0$, $\rm D^+$, and $\rm D^{*+}$  cross sections were measured, 
though with limited precision and $\pt$ coverage, in 
 pp collisions at $\sqrt{s}=2.76~\tev$ using a sample of about $6\times 10^7$ minimum-bias events
 collected during a short run at the 
same energy as Pb--Pb collisions. These measurements  
were found to be in 
agreement with the scaled $7~\tev$ measurements,
within statistical uncertainties of about 20--40\% depending on $\pt$ and on the meson species~\cite{Dpp2.76}.

\section{Systematic uncertainties}
\label{sec:systematics}
{\it Systematic uncertainties on the {\rm Pb--Pb} yields}

The systematic uncertainties on the prompt D meson yields in Pb--Pb collisions
are summarized in Table~\ref{tab:SystPbPbSpectra} for the lowest and highest $\pt$ intervals in the two centrality classes 0--20\% and 40--80\%. 

The systematic uncertainty on the yield extraction from the invariant 
mass spectra was determined by repeating the fit, in each $\pt$ interval, 
in a different mass range and also with a different function to describe the background. 
Namely, a parabola, instead of an exponential, was considered for $\Dzero$ and $\Dplus$, and 
a power law multiplied by an exponential or a polynomial for $\Dstar$. 
A method based on counting the signal in the invariant mass distribution, 
after subtraction of the background estimated from a fit to the side bands,
was also used. 
The uncertainty was defined as the maximum difference of these results and it 
was found to vary 
in the range 5--20\%, depending on the $\pt$ interval and on the collision centrality.

The systematic uncertainty on the tracking efficiency
was estimated by comparing  
the efficiency (i) of track finding in the TPC and (ii) of track prolongation from the 
TPC to the ITS between data and simulation, and (iii) by varying the track quality selections.
The efficiency of track prolongation from the TPC to the ITS  and of association of hits in the 
silicon pixel layers was found to be described in simulation
at the level of 5\% in the $\pt$ range relevant for this analysis 
(0.5--15~GeV/$c$). The centrality dependence of these efficiencies, which is limited 
to $\pm 3\%$ in this $\pt$ range, was found to be reproduced within 1.5\%. 
The effect of wrong association of ITS hits to tracks was studied in the simulation.
It was found that the fraction of D mesons with at least one
decay track with wrong hit associations increases with centrality, due to the
higher detector occupancy, and vanishes at large $\pt$, where the 
track extrapolation between layers is more precise. 
In the centrality class 0--20\%, it ranges from 7\% to 1\% in the 
transverse momentum interval $2<\pt<16~\gev/c$.
However, it was verified that the signal selection efficiencies
are compatible, within statistical uncertainties, between D mesons with and 
without wrong hit associations. Indeed, the mis-associated hit is typically
very close in space to the correct hit. 
Overall, the systematic uncertainty from track reconstruction amounts 
to 5\% for single tracks, which
results in a 10\% uncertainty for $\Dzero$ mesons (two-track final state) and 15\% uncertainty 
for $\Dplus$ and $\Dstar$ mesons (three-track final state).

\begin{table}[!t]
\caption{Summary of relative systematic uncertainties on the prompt D meson production yields in Pb--Pb collisions for the lowest and highest $\pt$ 
bins measured for the three mesons.}
\centering
\renewcommand{\arraystretch}{1.3}
\begin{tabular}{|c|l|cc|cc|cc|cc|}
\hline 
\multicolumn{2}{|r|}{Particle} & \multicolumn{2}{c|}{$\Dzero$} 
 & \multicolumn{2}{c|}{$\Dplus$}& \multicolumn{2}{c|}{$\Dstar$} \\
\hline
\multirow{9}{*}{\parbox{2 cm}{\centering 0--20\%\\ centrality}} & 
\multicolumn{1}{r|}{$\pt$ interval ($\gev/c$)} 
& 2--3 & 12--16 & 6--8 & 12--16 & 4--6 & 12--16\\
\cline{2-8} 
& Yield extraction      & \phantom{0}8\%	& 10\%
                      & 20\% & 10\%   
                      & 20\% & 10\%   \\
& Tracking efficiency & 10\% & 10\%
                      & 15\% & 15\%  
                      & 15\% & 15\%  \\
& Cut efficiency      &   13\% & 10\% 
                      &   15\% & 15\%  
                      &   10\% & 10\%  \\
& PID efficiency      &   $_{-\phantom{0}5} ^{+15}\%$ & \phantom{0}5\%
                      &   \phantom{0}5\% & \phantom{0}5\%
                      &   $_{-\phantom{0}5} ^{+15}\%$ & \phantom{0}5\%\\
& MC $\pt$ shape      & \phantom{0}4\% & \phantom{0}3\%  
                      & \phantom{0}1\% & \phantom{0}5\%  
                      & \phantom{0}3\% & \phantom{0}3\% \\
& FONLL feed-down corr.   & $_{-14}^{+\phantom{0}2}\%$ & $_{-8}^{+6}\%$ 
                          & $_{-7}^{+3}\%$ & $_{-9}^{+7}\%$ 
                          & $_{-\phantom{0}5}^{+\phantom{0}2}\%$ & $_{-\phantom{0}7}^{+\phantom{0}2}\%$  \\ 
& $R_{{\rm AA}}^{\rm feed-down} / R_{{\rm AA}}^{\rm prompt}$ (Eq.~(\ref{eq:fcNbMethod}))
      		      & $^{+\phantom{0}4}_{-10} \%$  & $^{+14}_{-27} \%$
                      &	$^{+\phantom{0}7}_{-16} \%$  & $^{+15}_{-28} \%$ 
                      &	$^{+\phantom{0}4}_{-\phantom{0}9} \%$  & $^{+\phantom{0}5}_{-12} \%$ \\
& BR         & \multicolumn{2}{c|}{ 1.3\%} & \multicolumn{2}{c|}{2.1\%}& \multicolumn{2}{c|}{ 1.5\%}\\
\cline{3-8}
& Centrality limits & \multicolumn{6}{c|}{$<0.1$\%}\\
\hline
\multirow{9}{*}{\parbox{2 cm}{\centering 40--80\%\\ centrality}} & 
\multicolumn{1}{r|}{$\pt$ interval ($\gev/c$)} 
& 2--3 & 12--16 & 3--4 & 8--12 & 2--4 & 12--16\\
\cline{2-8}
& Yield extraction      & \phantom{0}5\%	& \phantom{0}5\%
                      & 15\%	& 15\%
                      & 15\%    & \phantom{0}8\%\\   
& Tracking efficiency & 10\% & 10\%
                      & 15\% & 15\%  
                      & 15\% & 15\%  \\
& Cut efficiency      &   13\% & 10\% 
                      &   10\% & 10\%  
                      &   10\% & 10\%  \\
& PID efficiency      &   $_{-\phantom{0}5} ^{+10}\%$ & \phantom{0}5\%
                      &   \phantom{0}5\% & \phantom{0}5\%
                      &   $_{-\phantom{0}5} ^{+10}\%$ & \phantom{0}5\%\\
& MC $\pt$ shape      & \phantom{0}1\% & \phantom{0}3\%  
                      & \phantom{0}1\% & \phantom{0}3\% 
                      & \phantom{0}5\% & \phantom{0}4\% \\
& FONLL feed-down corr.  & $_{-16}^{+\phantom{0}3}\%$ & $_{-\phantom{0}5}^{+\phantom{0}4}\%$ 
                      & $_{-11}^{+\phantom{0}3}\%$ & $_{-\phantom{0}9}^{+\phantom{0}4}\%$ 
                      & $_{-\phantom{0}8}^{+\phantom{0}1}\%$ & $_{-\phantom{0}4}^{+\phantom{0}1}\%$  \\
& $R_{{\rm AA}}^{\rm feed-down} / R_{{\rm AA}}^{\rm prompt}$ (Eq.~(\ref{eq:fcNbMethod}))	
                      &  $^{+\phantom{0}5}_{-12} \%$ & $^{+11}_{-22} \%$ 
                      &	  $^{+\phantom{0}6}_{-14} \%$  &  $^{+\phantom{0}9}_{-20} \%$ 
                      &	  $^{+\phantom{0}2}_{-\phantom{0}6} \%$  &  $^{+\phantom{0}3}_{-\phantom{0}8} \%$ \\
& BR         & \multicolumn{2}{c|}{1.3\%} & \multicolumn{2}{c|}{2.1\%}& \multicolumn{2}{c|}{1.5\%}\\
\cline{3-8}
& Centrality limits & \multicolumn{6}{c|}{3\%}\\
\hline
\end{tabular}
\label{tab:SystPbPbSpectra}
\end{table}

The uncertainty on the correction for the selection cut efficiency was evaluated by repeating the analysis with different sets of cuts and was defined as the variation 
of the resulting corrected yields about the value corresponding to the central set. 
This resulted in 13\% for $\Dzero$ for $\pt<3~\gev/c$, 15\% for 
$\Dplus$ in all $\pt$ intervals in the 0--20\% centrality class, and 10\% for the other cases (see Table~\ref{tab:SystPbPbSpectra}).
Part of this uncertainty comes from residual detector misalignment effects not 
fully described in the simulation. 
In order to estimate this contribution, the secondary vertices in the simulation were 
also reconstructed after scaling, for each track, the impact parameter 
residuals with respect to their true value.
In particular, a scaling factor of 1.1--1.2 was applied in order to
reproduce the impact parameter resolution observed in the data 
(see Fig.~\ref{fig:imppar}).
The relative variation of the efficiency is 8\% for 
$\pt = 2$--$3~\gev/c$ and negligible for $\pt > 5~\gev/c$. 
This effect was not included explicitly in the systematic uncertainty, 
since it is already accounted for in the cut variation study.
A further check was performed by comparing the distributions
of the cut variables used for the candidate selection
 in the data 
and in the simulation.
These comparisons can only be carried out by releasing the selection, hence essentially for background
candidates. However, they provide an indication of the level of accuracy of the simulation.
A good agreement was observed, with no dependence on collision centrality.

The uncertainty arising from the PID selection was estimated by
comparing the corrected signals extracted with and without this selection.
In the 20\% most central collisions, it was found to be 
$_{-\phantom{0}5\%} ^{+15\%}$ for $\pt<6~\gev/c$ and $\pm 5\%$ for $\pt>6~\gev/c$.
In the 40--80\% centrality range, it was estimated to be $\pm 5\%$ for
$\pt>3~\gev/c$ and $_{-\phantom{0}5\%} ^{+10\%}$ in $2<\pt<3~\gev/c$.

The uncertainty on the efficiencies arising from the difference
between the real and simulated shape of the D meson transverse momentum 
distribution, which includes also the effect of the $\pt$ dependence of the 
nuclear modification factor, depends on the width of the $\pt$ intervals and on 
the slope of the efficiencies with $\pt$.
It was estimated by varying the simulated shape between the 
PYTHIA and FONLL ${\rm d}N/{\rm d}\pt$, with and without the nuclear 
modification observed in the data. 
The resulting uncertainty is below 5\% in all the $\pt$ intervals considered 
for the three meson species.
As an example, for $\Dzero$ it is 4\% in the lowest and highest $\pt$ 
intervals (2--3~$\gev/c$ and 12--16~$\gev/c$) and 1\% in 3--12~$\gev/c$.

The $\pt$-differential yields for $\Dzero$ and $\overline{\rm D}^0$ mesons, extracted separately,
were found to be in agreement
within the statistical uncertainties of about 20--25\%. Due to the limited statistics, this check could not
be carried out for $\Dplus$ and $\Dstar$ mesons.

\begin{figure}[!b]
  \begin{center}
    \includegraphics[width=0.49\textwidth]{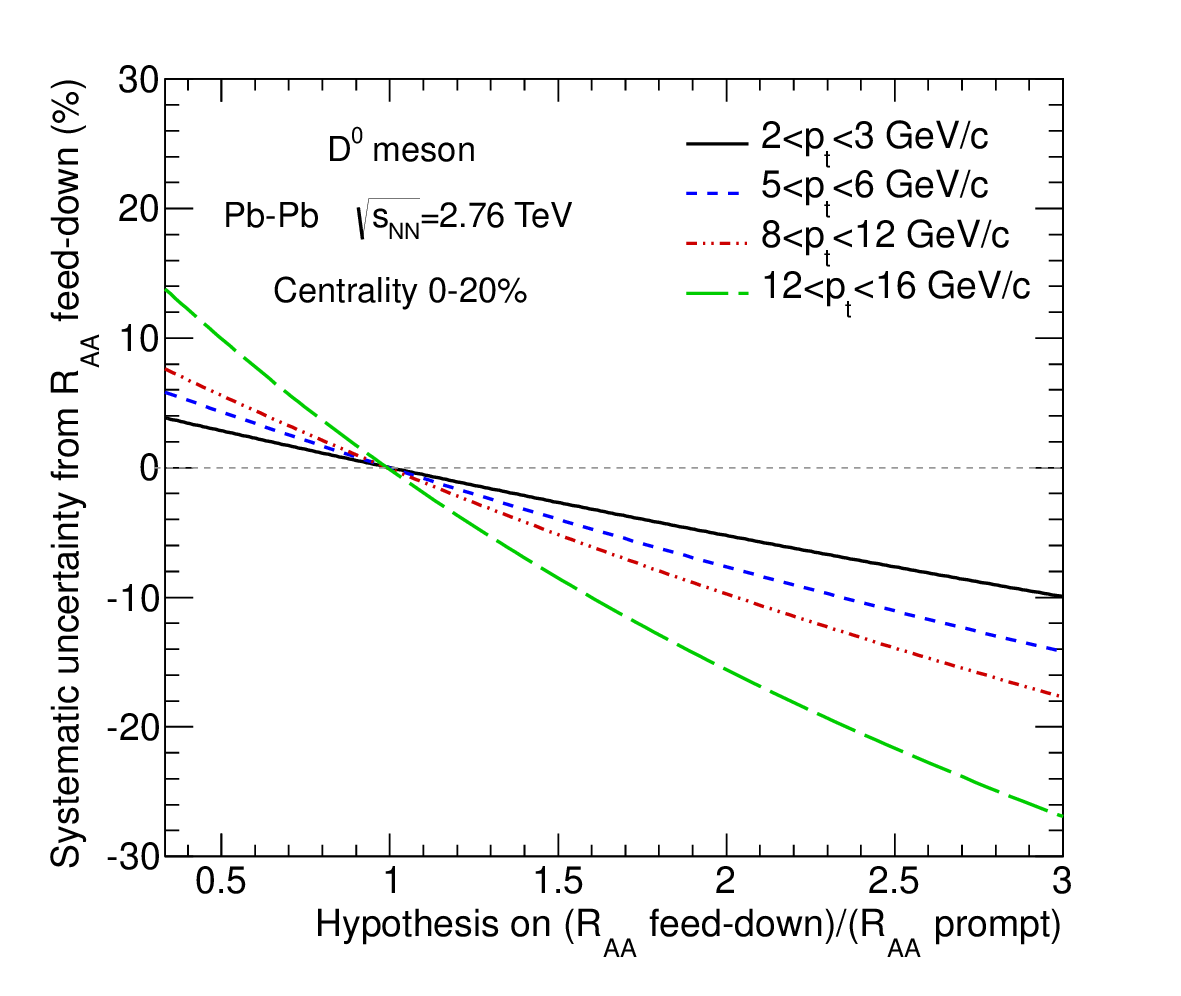}
\caption{Relative variation of the prompt $\Dzero$ meson yield as a function of the hypothesis on  $\RAA^{\rm feed-down}/\RAA^{\rm prompt}$ for the B feed-down subtraction approach based on Eq.~(\ref{eq:fcNbMethod}).}
\label{fig:Raab_vs_Raac}
\end{center}
\end{figure}

The systematic uncertainty from the subtraction of feed-down D mesons from B 
decays was estimated as for the pp case~\cite{Dpp7paper}. 
The contribution of the FONLL 
perturbative uncertainties was included by varying the heavy quark masses and the 
factorization and renormalization scales in the ranges proposed
in~\cite{FONLLrhic}.
Furthermore, a different procedure was used to evaluate 
the prompt fraction. In this approach,
the ratio of the FONLL feed-down and prompt 
production cross sections is the input for evaluating the correction factor. 
Then, the prompt fraction depends explicitly on
the ratio of nuclear modification factors of feed-down and prompt D mesons:
\begin{equation}
f^\prime_{\rm prompt} = \left( 	1 + 	\frac{({\rm Acc}\times\epsilon)_{{\rm feed-down}}}{({\rm Acc}\times\epsilon)_{{\rm prompt}}}	\cdot
		 \frac{ \left(\frac{{\rm d}^2 \sigma}{{\rm d}y \, {\rm d} \pt } \right)^{{\sf FONLL}}_{{\rm feed-down}} }{ \left(\frac{{\rm d}^2 \sigma}{{\rm d}y \, {\rm d} \pt } \right)^{{\sf FONLL}}_{ {\rm prompt} } } 
		 \cdot \frac{ \RAA^{\rm feed-down} } { \RAA^{\rm prompt} }
\right)^{-1} \, .
\label{eq:fc}
\end{equation}

The systematic uncertainty due to the B feed-down subtraction was 
evaluated as the envelope of the results obtained with the two methods, 
Eqs.~(\ref{eq:fcNbMethod}) and~(\ref{eq:fc}),
when varying the FONLL parameters.
The resulting uncertainty ranges between $^{+\phantom{0}2}_{-14}\%$ 
at low $\pt$ and $^{+6}_{-8}\%$ at high $\pt$ (for $\Dzero$ in the 0--20\% centrality class).

The contribution from the different nuclear modification factors of prompt and
feed-down D mesons was evaluated by varying the hypothesis on 
$\RAA^{\rm feed-down}/\RAA^{\rm prompt}$ in the range $1/3<\RAA^{\rm feed-down}/\RAA^{\rm prompt}<3$ for both feed-down subtraction methods.
The resulting uncertainty is at most 30\%, as shown in Fig.~\ref{fig:Raab_vs_Raac}, 
where the relative prompt $\Dzero$ yield variation 
is displayed as a function of $\RAA^{\rm feed-down}/\RAA^{\rm prompt}$ for four
$\pt$ intervals using the B feed-down subtraction approach 
based on Eq.~(\ref{eq:fcNbMethod}). 
Considering the resulting values of $\RAA^{\rm prompt}$ shown in the next section,
the variation of the hypothesis on $\RAA^{\rm feed-down}/\RAA^{\rm prompt}$ 
corresponds, for the 20\% most central collisions, to values of the nuclear 
modification factor of D mesons from 
B feed-down in a range of about 0.17--1.5 at low $\pt$ and 0.09--0.8 
at high $\pt$.
The $\RAA$ of non-prompt $\rm J/\psi$, measured by CMS~\cite{CMSquarkonia}, 
falls in this range as well as the available model predictions for 
B meson energy loss~\cite{whdg,adsw}.

The contribution due to the 1.1\% relative uncertainty on the fraction of 
hadronic cross section used in the Glauber fit to determine the centrality 
classes was obtained by estimating the variation of the D meson 
$\d N/\d\pt$ 
when the limits of the centrality classes are shifted by $\pm$1.1\% 
(e.g., for the 40--80\% class, 40.4--80.9\% and 39.6--79.1\%). 
The resulting uncertainty is common to all meson species and all $\pt$ bins for 
a given centrality class. It increases from central to peripheral events. 
In particular, it is less than 0.1\% in the 0--20\% centrality class and 
3\% in 40--80\%.

Finally, the systematic uncertainty on the branching ratios~\cite{pdg}
was considered.

{\it Systematic uncertainties on $\RAA$}

\begin{table}[!t]
\caption{Summary of relative systematic uncertainties on $R_{{\rm AA}}$. 
For the data systematic uncertainties and the B feed-down subtraction some
of the contributions are singled-out in the indented rows.}
\centering
\renewcommand{\arraystretch}{1.3}
\begin{tabular}{|c|l|cc|cc|cc|}
\hline 
\multicolumn{2}{|r|}{Particle} & \multicolumn{2}{c|}{$\Dzero$} 
 & \multicolumn{2}{c|}{$\Dplus$}& \multicolumn{2}{c|}{$\Dstar$} \\
\hline
\multirow{9}{*}{\parbox{2 cm}{\centering 0--20\%\\ centrality}} & 
\multicolumn{1}{r|}{$\pt$ interval ($\gev/c$)} 
& 2--3 & 12--16 & 6--8 & 12--16 & 4--6 & 12--16\\
\cline{2-8} 
& Data syst. pp and Pb--Pb     
                      &	$^{+33}_{-41} \%$  & $^{+28}_{-28} \%$
                      &	$^{+35}_{-35} \%$  & $^{+35}_{-35} \%$   
                      &	$^{+42}_{-41} \%$  & $^{+34}_{-35} \%$   \\
&  {\small $\qquad$ Data syst. in Pb--Pb }      
                      &	$^{+26}_{-22} \%$  & $^{+22}_{-22} \%$ 
                      &	$^{+30}_{-30} \%$  & $^{+27}_{-27} \%$
                      &	$^{+39}_{-36} \%$  & $^{+29}_{-29} \%$ \\
&  {\small $\qquad$ Data syst. in pp }      
                      &	17\%             & 17\% 
                      &	15\%             & 21\%
                      &	15\%             & 18\% \\
&  {\small $\qquad$ $\sqrt{s}$-scaling of the pp ref. }      
                      &	$^{+10}_{-31} \%$  & $^{+\phantom{0}5}_{-\phantom{0}6} \%$ 
                      &	$^{+\phantom{0}6}_{-10} \%$  & $^{+\phantom{0}4}_{-\phantom{0}6} \%$
                      &	$^{+\phantom{0}7}_{-14} \%$  & $^{+\phantom{0}5}_{-\phantom{0}6} \%$ \\
\cline{2-8}
& Feed-down subtraction  &	$^{+15}_{-14} \%$ & $^{+16}_{-29} \%$
                      &	 $^{+12}_{-18} \%$ & $^{+17}_{-28} \%$
                      &	$^{+\phantom{0}5}_{-12} \%$  & $^{+\phantom{0}8}_{-16} \%$ \\
& {\small $\qquad$ FONLL feed-down corr.  }   
		   &	$^{+12}_{-\phantom{0}2} \%$ & $^{+\phantom{0}1}_{-\phantom{0}2} \%$
                      &	 $^{+\phantom{0}3}_{-\phantom{0}2} \%$ & $^{+\phantom{0}2}_{-\phantom{0}1} \%$
                      &	$^{+\phantom{0}1}_{-\phantom{0}1} \%$  & $^{+\phantom{0}2}_{-\phantom{0}1} \%$ \\
& {\small $\qquad$ $R_{{\rm AA}}^{\rm feed-down} / R_{{\rm AA}}^{\rm prompt}$ (Eq.~(\ref{eq:fcNbMethod}))}	 
		  & $^{+\phantom{0}4}_{-10} \%$ &	 $^{+14}_{-27} \%$
                      &	 $^{+\phantom{0}7}_{-16} \%$  & $^{+15}_{-28} \%$ 
                      &	 $^{+\phantom{0}4}_{-\phantom{0}9} \%$  & $^{+\phantom{0}5}_{-12} \%$ \\
\cline{2-8}
& Normalization         & \multicolumn{6}{c|}{5.3\%}\\
\hline 
\multirow{9}{*}{\parbox{2 cm}{\centering 40--80\%\\ centrality}} & 
\multicolumn{1}{r|}{$\pt$ interval ($\gev/c$)} 
& 2--3 & 12--16 & 3--4 & 8--12 & 2--4 & 12--16\\
\cline{2-8}
& Data syst. pp and Pb--Pb     
                      &	  $^{+28}_{-40} \%$  &  $^{+24}_{-25} \%$
                      &	 $^{+40}_{-43}\%$  & $^{+30}_{-31} \%$
                      &	$^{+33}_{-39} \%$  & $^{+29}_{-30}\%$   \\
&  {\small $\qquad$ Data syst. in Pb--Pb }      
                      &	$^{+21}_{-19} \%$  & $^{+17}_{-17} \%$ 
                      &	$^{+25}_{-25} \%$  & $^{+24}_{-24} \%$
                      &	$^{+28}_{-27} \%$  & $^{+22}_{-22} \%$ \\
&  {\small $\qquad$ Data syst. in pp }      
                      &	17\%  & 17\% 
                      &	30\%  & 17\%
                      &	15\%  & 18\% \\
&  {\small $\qquad$ $\sqrt{s}$-scaling of the pp ref.   }  
                      &	$^{+10}_{-31} \%$  & $^{+\phantom{0}5}_{-\phantom{0}6} \%$ 
                      &	 $^{+\phantom{0}8}_{-19} \%$ & $^{+\phantom{0}5}_{-\phantom{0}8} \%$ 
                      &	$^{+10}_{-24} \%$& $^{+\phantom{0}5}_{-\phantom{0}6} \%$   \\
\cline{2-8}
& Feed-down subtraction  &	$^{+13}_{-17} \%$ & $^{+12}_{-23} \%$
                      &	 $^{+10}_{-18} \%$ & $^{+15}_{-25} \%$
                      &	$^{+\phantom{0}3}_{-13} \%$  & $^{+\phantom{0}3}_{-14} \%$ \\
&  {\small $\qquad$ FONLL feed-down corr. }    
		   &	 $^{+10}_{-\phantom{0}2} \%$  & $^{+\phantom{0}1}_{-\phantom{0}1} \%$
                      &	 $^{+\phantom{0}4}_{-\phantom{0}1} \%$  & $^{+\phantom{0}2}_{-\phantom{0}1} \%$
                       &	  $^{+\phantom{0}1}_{-\phantom{0}5} \%$ &  $^{+\phantom{0}1}_{-\phantom{0}3} \%$ \\
&  {\small $\qquad$ $R_{{\rm AA}}^{\rm feed-down} / R_{{\rm AA}}^{\rm prompt}$ (Eq.~(\ref{eq:fcNbMethod}))}	
		   &  $^{+\phantom{0}5}_{-12} \%$ & $^{+11}_{-22} \%$ 
                      &	  $^{+\phantom{0}6}_{-14} \%$  &  $^{+9}_{-20} \%$ 
                      &	  $^{+\phantom{0}2}_{-\phantom{0}6} \%$  &  $^{+\phantom{0}3}_{-\phantom{0}8} \%$ \\
\cline{2-8}
& Normalization         & \multicolumn{6}{c|}{7.5\%}\\
\hline
\end{tabular}
\label{tab:SystRaa}
\end{table}

The systematic uncertainties on the $\Raa$ measurement 
derive from the uncertainties on: 
the reference cross section for pp collisions, the Pb--Pb yields, and the 
average 
nuclear overlap function for the various centrality classes, as given in Table~\ref{tab:centbins}. 
For the pp reference, the uncertainties on the measurement at $\sqrt{s}=7~\TeV$ were 
quantified in~\cite{Dpp7paper} and the scaling to $\sqrt{s}=2.76~\TeV$, 
described in Section~\ref{sec:reference}, introduces additional 
uncertainties of about 10--30\%.
The uncertainties on the Pb--Pb prompt D meson yields were described previously in this section.
For the nuclear modification factor, 
the pp and Pb--Pb uncertainties were added in quadrature, except for 
the feed-down contribution deriving from FONLL uncertainties,
that partly cancels in the ratio. 
This contribution was evaluated by comparing the $\RAA$ values obtained
with the two methods for feed-down correction described above and
with the different heavy quark masses, factorization and renormalization 
scales used in FONLL.
In this study, the same method and the same set of FONLL parameters were used
for pp and Pb--Pb, so as to take into account the correlations of these sources 
in the numerator and denominator of $\RAA$.

 
The resulting systematic uncertainties are summarized in 
Table~\ref{tab:SystRaa}. 
In the table, the normalization uncertainty is the quadratic sum of the 3.5\% 
pp normalization uncertainty~\cite{Dpp7paper}, the contribution due to the 1.1\% 
uncertainty on the fraction of hadronic cross section used in the Glauber fit 
discussed above, and the uncertainty on $\av{\TAA}$, which is 3.9\% for the 
centrality class 0--20\% and 5.9\% for the 40--80\% class.

\section{Results}
\label{sec:results}
\subsection{D meson $\pt$ spectra and $\RAA$}

\begin{figure}[!t]
\begin{center}
\includegraphics[width=\textwidth]{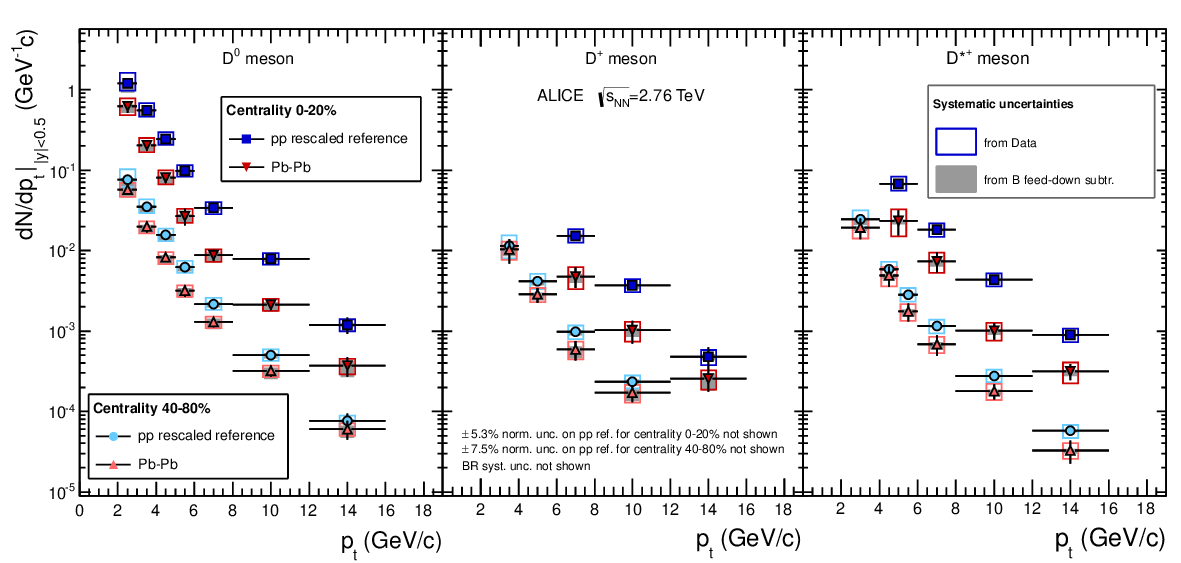}
\caption{(colour online) Transverse momentum distributions $\d N/\d\pt$ of 
prompt $\rm D^0$ (left) and $\rm D^+$ (centre), and  $\rm D^{*+}$ (right) 
mesons in the 0--20\% and 40--80\% centrality classes in Pb--Pb collisions 
at $\sqrtsNN=2.76~\tev$. 
The reference pp distributions $\av{\TAA}\,\d \sigma/\d\pt$ are shown as well. 
Statistical uncertainties (bars) and systematic uncertainties from data 
analysis (empty boxes) and from feed-down subtraction (full boxes) are shown. 
For Pb--Pb, the latter includes the uncertainties from the FONLL feed-down 
correction and from the variation of the hypothesis on
$R_{{\rm AA}}^{\rm prompt} / R_{{\rm AA}}^{\rm feed-down}$.
Horizontal error bars reflect bin widths, symbols were placed at the centre of the bin.}
\label{fig:spectra}
\end{center}
\end{figure}

The transverse momentum distributions $\d N/\d\pt$ 
of prompt $\rm D^0$, $\rm D^+$, and $\rm D^{*+}$ mesons
are presented in Fig.~\ref{fig:spectra},
for the centrality classes 0--20\% and 40--80\%. 
The spectra from Pb--Pb
collisions, defined as the feed-down corrected production yields per event (see
Eq.~(\ref{eq:dNdpt})), are compared to the reference spectra from pp
collisions, which are constructed as $\av{\TAA}\,\d\sigma/\d\pt$, 
using the $\sqrt{s}$-scaled pp measurements at 7~TeV~\cite{Dpp7paper}
and the average nuclear overlap function values from Table~\ref{tab:centbins}. 
A clear suppression is observed in Pb--Pb collisions, which is stronger in central than in peripheral collisions.

\begin{figure}[!t]
  \begin{center}
  \includegraphics[width=\textwidth]{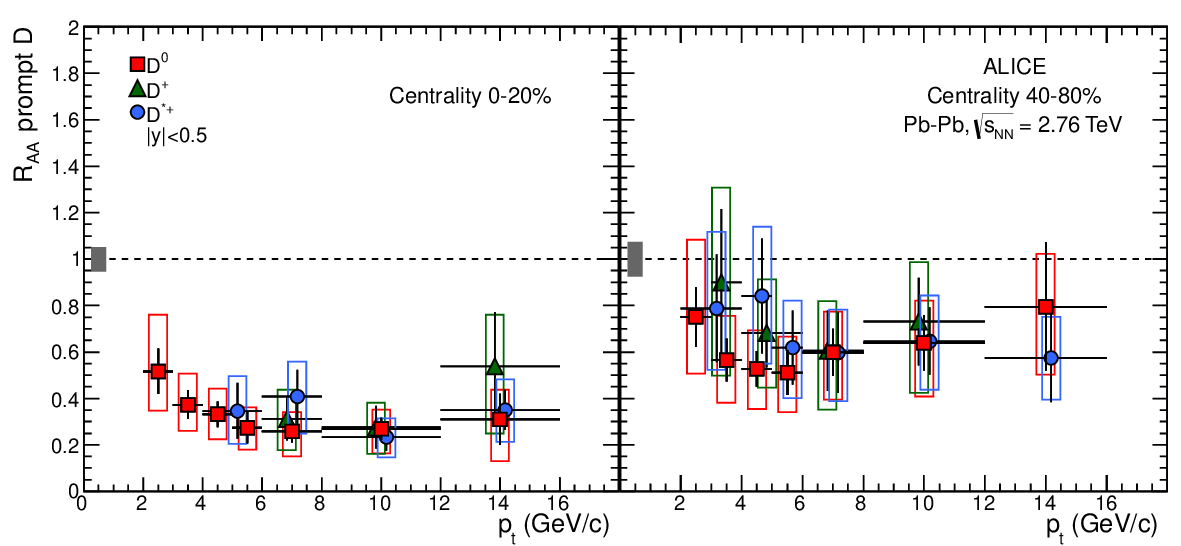}
  \caption{(colour online) $\RAA$ 
  for prompt $\rm D^0$, $\rm D^+$, and $\Dstar$ in the 0--20\% (left) and 40--80\% (right) centrality classes. Statistical (bars), 
  systematic (empty boxes), and normalization (full box) uncertainties are shown.
    Horizontal error bars reflect bin widths, symbols were placed at the centre of the bin.}
\label{fig:RAApt}
\end{center}
\end{figure}

The ratio of the Pb--Pb to the reference spectra provides the
nuclear modification factors $\RAA(\pt)$ 
of prompt $\rm D^0$, $\rm D^+$, and $\Dstar$ mesons, which are shown 
for central 
(0--20\%) and semi-peripheral (40--80\%) collisions in Fig.~\ref{fig:RAApt}. 
The vertical bars represent the statistical uncertainties, typically about 20--25\%
for $\Dzero$  and about 30--40\% for $\Dplus$ and $\Dstar$ mesons in central collisions. The total 
$\pt$-dependent systematic uncertainties, shown as empty boxes, include all the contributions 
described in the previous section, except for the normalization uncertainty, which 
is displayed as a filled box at $\RAA=1$. 
The results for the three D meson species are in agreement within statistical uncertainties and they 
show a
 suppression reaching a factor 3--4 ($\RAA\approx 0.25$--0.3) in central collisions for $\pt>5~\gev/c$.
For decreasing $\pt$, the $\Dzero$ 
$\RAA$ in central collisions shows a tendency to less suppression.

\begin{figure}[!t]
  \begin{center}
\includegraphics[width=\textwidth]{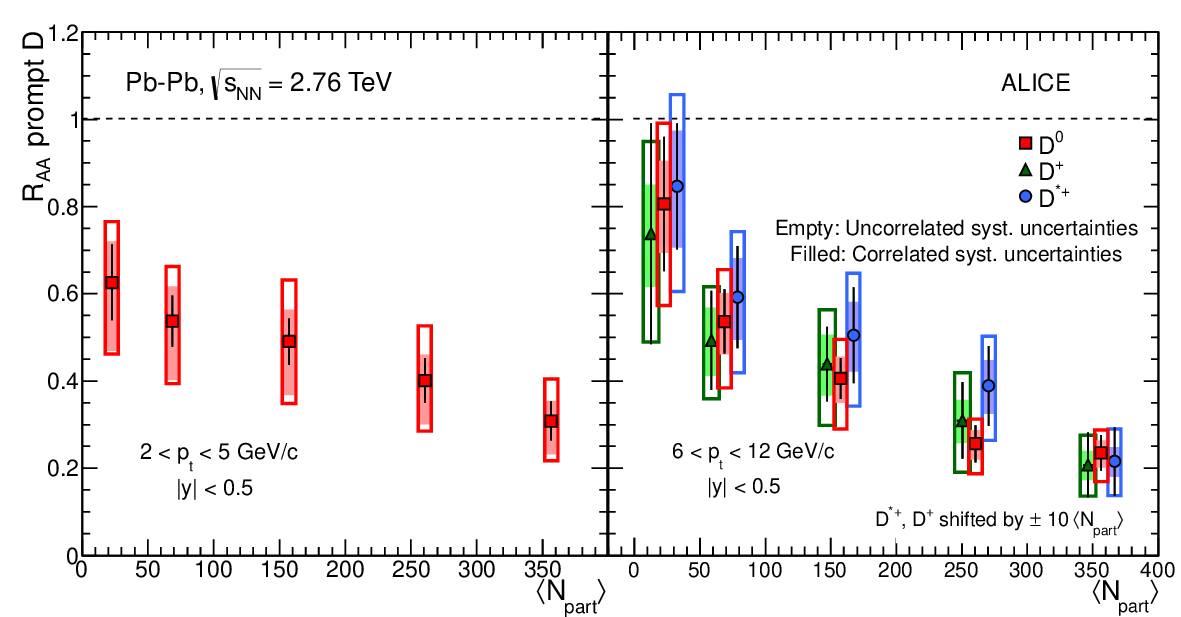}
\caption{Centrality dependence of $\RAA$ for prompt D mesons. 
Left: $\Dzero$ mesons with $2<\pt<5~\gev/c$. 
Right: $\Dzero$, $\Dplus$, and $\Dstar$ mesons with $6<\pt<12~\gev/c$.
$\Dplus$ and $\Dstar$ points are displaced horizontally for better visibility.}
\label{fig:RAAcentr}
\end{center}
\end{figure}

The centrality dependence of the nuclear modification factor was studied
in the two wider transverse momentum intervals $2<\pt<5~\gev/c$, 
for $\Dzero$, and $6<\pt<12~\gev/c$, for the three D meson species.
This study was performed in five centrality classes from 0--10\% to 60--80\% 
(see Table~\ref{tab:centbins}). 
The invariant mass analysis and all the corrections were carried out 
as described in Sections~\ref{sec:signal} and~\ref{sec:corrections}.
The systematic uncertainties are essentially the same as for the 
$\pt$-dependence analysis, except for the contribution from the D meson 
$\pt$-shape in the simulation, which is larger in the wide intervals. 
It amounts to 8\% for $\Dzero$, 10\% for $\Dplus$, and 
5--15\% (depending on centrality) for $\Dstar$ mesons in $6<\pt<12~\gev/c$.
In the transverse momentum interval 2--5~$\gev/c$, this uncertainty is larger
(8--17\%, depending on centrality) due to the larger contribution from the 
$\pt$ dependence of the nuclear modification factor.
The resulting $\RAA$ is shown in Fig.~\ref{fig:RAAcentr} as a function
of the average number of participants, $\langle\Npart\rangle$.
The contribution to the systematic uncertainty that is fully correlated 
between centrality classes (normalization and pp reference cross-section) and 
the remaining, uncorrelated, systematic uncertainties are displayed 
separately, by the filled and empty boxes, respectively.
The contribution from feed-down correction was considered among the 
uncorrelated sources because it is dominated by the variation of the 
ratio $\RAA^{\rm feed-down}/\RAA^{\rm prompt}$, which may depend on
centrality.
For the $\pt$ interval 6--12~$\gev/c$, the suppression increases with
increasing centrality.
It is interesting to note that the suppression of prompt D mesons at central 
rapidity and high transverse momentum, shown in the right-hand panel of 
Fig.~\ref{fig:RAAcentr} is very similar, both in size and centrality 
dependence, to that of prompt J/$\psi$ mesons in a similar $\pt$ range and 
$|y|<2.4$, recently measured by the CMS Collaboration~\cite{CMSquarkonia}.

\subsection{Comparisons to light-flavour hadrons and with models}
\label{sec:comparisons}

In this section, the average nuclear modification factor of the three D meson species is compared to 
that of charged particles~\cite{chargedRAA}, mainly light-flavour hadrons, and to model calculations. The contributions of 
$\Dzero$, $\Dplus$, and $\Dstar$ to the average were weighted by their statistical uncertainties. Therefore,
the resulting $\RAA$ is close to that of the $\Dzero$~meson, which has the 
smallest uncertainties.
The systematic errors were calculated by propagating the uncertainties through
the weighted average, where the contributions from the tracking efficiency, 
from the B feed-down correction, and from the FONLL scaling of 7 TeV data to 
2.76 TeV were taken as fully correlated among the three D meson species.
The possible statistical correlation between the $\Dzero$ and $\Dstar$ $\RAA$, 
induced by the $\DstartoDpi$ decay, is negligible, because the 
statistical uncertainties, used as weights, are mainly determined by the 
background uncertainties, which are uncorrelated.
The resulting values are shown in Table~\ref{tab:averD1} 
for the two centrality classes where $\RAA$ was
measured as a function of $\pt$, and in Table~\ref{tab:averD2}  
for the $\RAA$ as a function of centrality in the transverse momentum 
range $6<\pt<12~\gev/c$.

\begin{table}[!t]
\caption{Average $\RAA$ as a function of $\pt$ for prompt D mesons in the 
0--20\% and 40--80\% centrality classes. The systematic error does
not include the normalization uncertainty, which is $\pm$5.3\%~($\pm$7.5\%) 
for the 0--20\%~(40--80\%) centrality class.}
\label{tab:averD1}
\centering
\renewcommand{\arraystretch}{1.3}
\begin{tabular}{|c|c|c|}
\hline 
$\pt$ interval & \multicolumn{2}{c|}{$\RAA~\pm$~stat~$\pm$~syst}\\
($\gev/c$) & 0--20\% centrality & 40--80\% centrality\\
\hline 
2--3 & $0.51\pm0.10_{-0.22}^{+0.18}$ & $0.75\pm0.13_{-0.32}^{+0.23}$\\
3--4 & $0.37\pm0.06_{-0.13}^{+0.11}$ & $0.59\pm0.09_{-0.21}^{+0.15}$\\
4--5 & $0.33\pm0.05_{-0.11}^{+0.10}$ & $0.55\pm0.07_{-0.18}^{+0.14}$\\
5--6 & $0.27\pm0.07_{-0.09}^{+0.08}$ & $0.54\pm0.08_{-0.17}^{+0.13}$\\
6--8 & $0.28\pm0.04_{-0.08}^{+0.07}$ & $0.60\pm0.08_{-0.18}^{+0.14}$\\
\phantom{0}8--12 & $0.26\pm0.03_{-0.07}^{+0.06}$ & $0.66\pm0.08_{-0.20}^{+0.16}$\\
12--16 & $0.35\pm0.06_{-0.12}^{+0.10}$ & $0.64\pm0.16_{-0.18}^{+0.16}$\\
\hline
\end{tabular}
\end{table}

\begin{table}[!t]
\caption{Average $\RAA$ as a function of centrality for prompt D mesons in the 
transverse momentum interval $6<\pt<12~\gev/c$.}
\label{tab:averD2}
\centering
\renewcommand{\arraystretch}{1.3}
\begin{tabular}{|c|c|}
\hline 
Centrality & $\RAA$$\pm$~stat~$\pm$~syst(uncorr)~$\pm$~syst(corr)\\
\hline
\phantom{0}0--10\% & $0.23\pm0.03~_{-0.06}^{+0.05}~_{-0.03}^{+0.03}$\\
10--20\% & $0.28\pm0.04~_{-0.07}^{+0.06}~_{-0.04}^{+0.03}$\\
20--40\% & $0.42\pm0.04~_{-0.11}^{+0.08}~_{-0.06}^{+0.05}$\\
40--60\% & $0.54\pm0.05~_{-0.13}^{+0.10}~_{-0.08}^{+0.07}$\\
60--80\%& $0.81\pm0.10~_{-0.21}^{+0.16}~_{-0.12}^{+0.11}$\\
\hline
\end{tabular}
\end{table}

\begin{figure}[!t]
  \begin{center}
\includegraphics[width=0.49\textwidth]{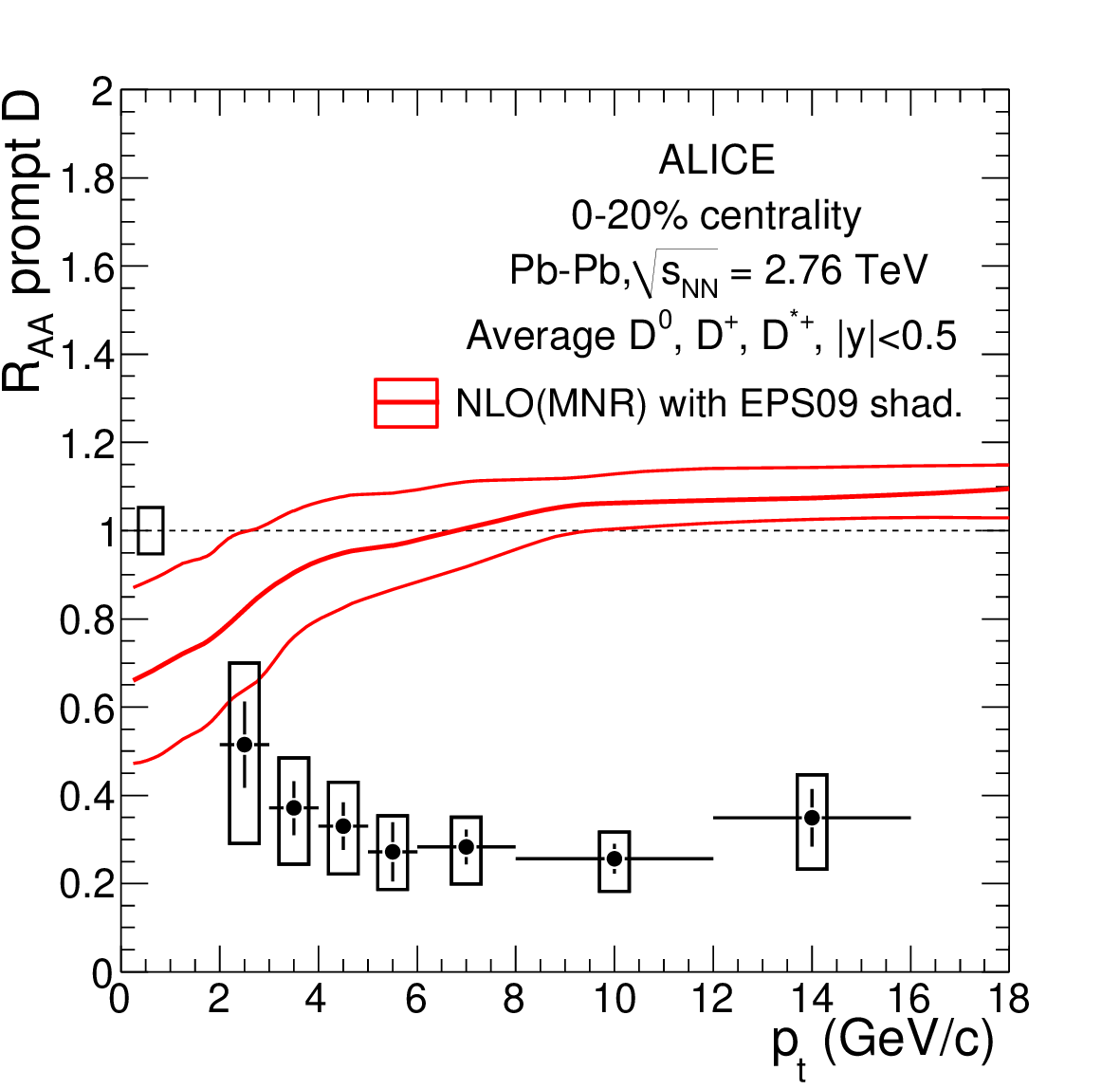}
\includegraphics[width=0.49\textwidth]{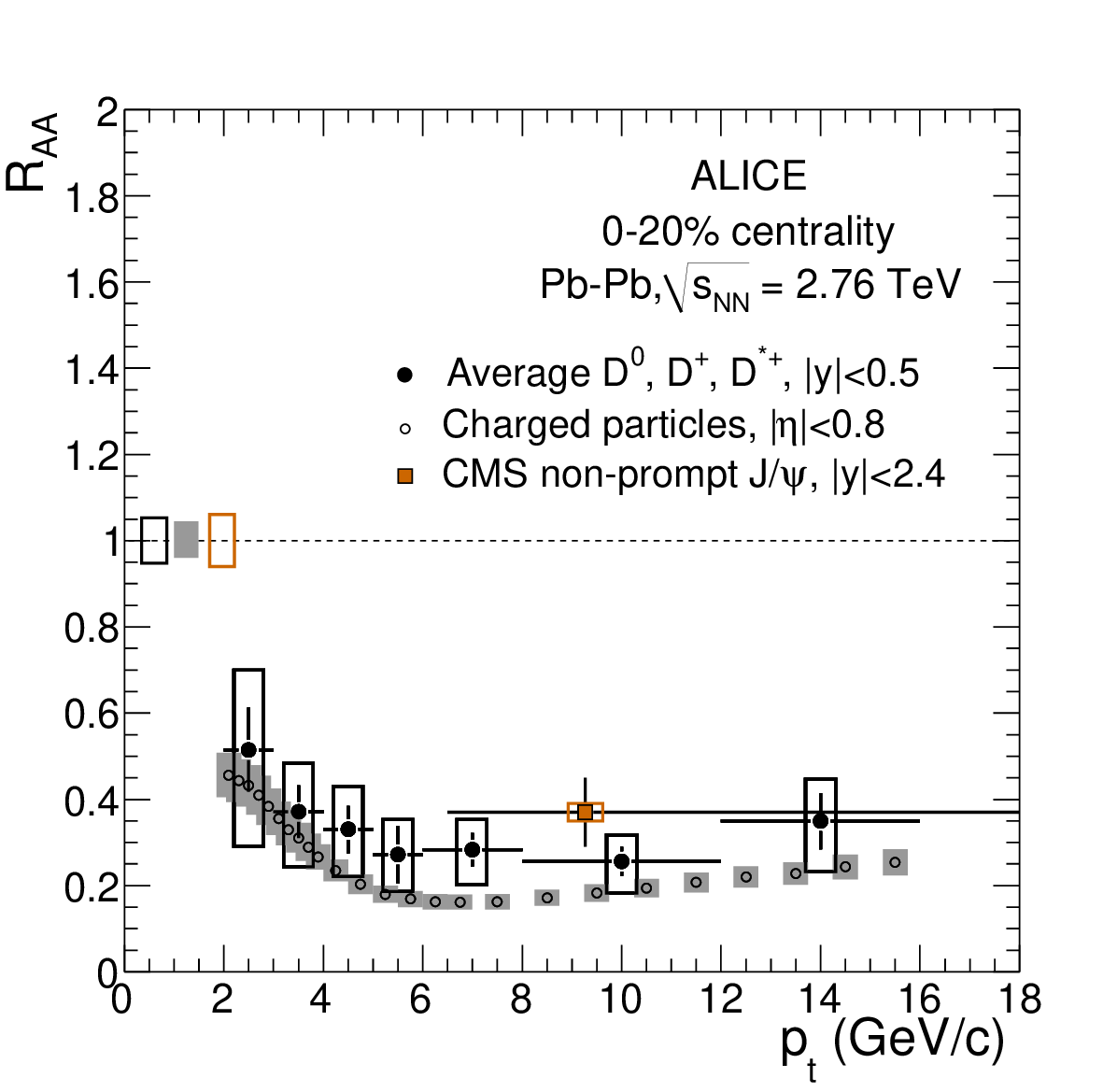}
  \caption{Average $\RAA$ of D mesons in the 0--20\% centrality class compared to: left, 
  the expectation from NLO pQCD~\cite{mnr} with nuclear shadowing~\cite{eps09}; right, 
  the nuclear modification factors of charged particles~\cite{chargedRAA} and non-prompt $\rm J/\psi$
  from B decays~\cite{CMSquarkonia} in the same centrality class. 
The charged particle $\RAA$ is shown only for \mbox{$2<\pt<16~\gev/c$}. The three normalization uncertainties shown 
in the right-hand panel are almost fully correlated.} 
\label{fig:RAApt_eps_charged}
\end{center}
\end{figure}

In addition to final state effects, where parton energy loss would be predominant, 
also initial-state effects are expected to influence the measured $\RAA$.
In particular, 
the nuclear modification of the parton distribution functions of the nucleons in the two colliding
nuclei modifies the initial hard scattering probability and, thus, the production yields of 
hard partons, including heavy quarks. In the kinematic range relevant for charm production 
at LHC energies, the main expected effect is nuclear shadowing, which reduces the parton distribution functions for partons with nucleon momentum fraction $x$ below $10^{-2}$.   
The effect of shadowing on the D meson $\RAA$ was estimated using the 
next-to-leading order (NLO) perturbative QCD calculation by Mangano, 
Nason, and Ridolfi (MNR)~\cite{mnr} with CTEQ6M parton distribution 
functions~\cite{cteq6} and the EPS09NLO parametrization~\cite{eps09} 
of their nuclear modification. The uncertainty band determined by the EPS09 uncertainties
is shown in the left-hand panel of Fig.~\ref{fig:RAApt_eps_charged}, together with the average D meson 
$\RAA$. The shadowing-induced effect on the $\RAA$
is limited to $\pm 15\%$ for $\pt>6~\gev/c$, suggesting that 
the strong suppression observed in the data is a final-state effect.

The expected colour charge and parton mass dependences of parton energy loss
should be addressed by comparing 
the nuclear modification factor of D and $\pi$ mesons. Since final results on the 
pion $\RAA$ at the LHC are not yet available, we compare here to 
charged particles.
Preliminary results~\cite{harryQM} have shown that the charged-pion $\RAA$
coincides with that of charged particles above $\pt\approx 5~\gev/c$
and it is lower by 30\% at $3~\gev/c$. 
The comparison between D meson and charged particle $\RAA$, reported in the 
right-hand panel of Fig.~\ref{fig:RAApt_eps_charged}, shows
that the average D meson nuclear modification factor is close to that of
charged particles~\cite{chargedRAA}.
However, considering that the systematic uncertainties of D mesons are not fully correlated with $\pt$, there is an indication for $\RAA^{\rm D}>\RAA^{\rm charged}$.
In the same figure, the nuclear modification factor measured by the CMS Collaboration for non-prompt 
J/$\psi$ mesons (from B decays) 
with $\pt>6.5~\gev/c$~\cite{CMSquarkonia} is also shown. Their suppression is clearly weaker
than that of charged particles, while the comparison with D mesons is not conclusive and 
would require more differential and precise measurements of the transverse momentum dependence.

\begin{figure}[!t]
  \begin{center}
\includegraphics[width=\textwidth]{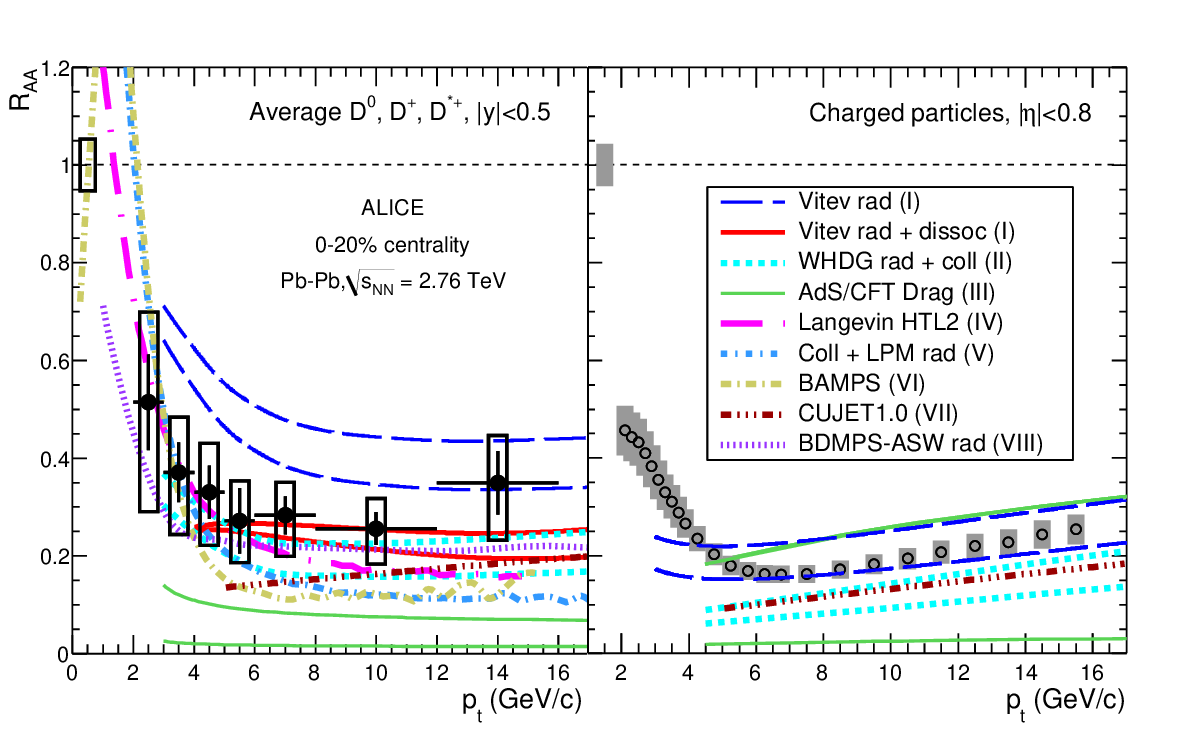}
  \caption{(colour online) Average $\RAA$ of D mesons (left) and $\RAA$ of charged particles (right)~\cite{chargedRAA} in the \mbox{0--20\%} centrality class compared to model calculations: (I)~\cite{vitev,vitevjet}, (II)~\cite{whdg2011}, (III)~\cite{horowitzAdSCFT}, (IV)~\cite{beraudo}, (V)~\cite{gossiaux}, (VI)~\cite{bamps}, (VII)~\cite{cujet}, (VIII)~\cite{adsw}. The two normalization uncertainties are almost fully correlated.}
\label{fig:RAApt_models}
\end{center}
\end{figure}

Several theoretical models based on parton energy loss compute the charm 
nuclear modification factor: (I)~\cite{vitev,vitevjet}, (II)~\cite{whdg2011}, (III)~\cite{horowitzAdSCFT}, (IV)~\cite{beraudo}, (V)~\cite{gossiaux}, (VI)~\cite{bamps}, (VII)~\cite{cujet}, (VIII)~\cite{adsw}. Figure~\ref{fig:RAApt_models} displays the comparison of these
models  
to the average D meson $\RAA$, for central Pb--Pb collisions (0--20\%), along with the 
comparison to the charged-particle $\RAA$~\cite{chargedRAA}, for those models that also compute 
this observable: (I)~\cite{vitev}, (II)~\cite{whdg2011}, (III)~\cite{horowitzAdSCFT}, (VII)~\cite{cujet}. Among the models that compute both observables, 
radiative energy loss supplemented with in-medium D meson dissociation (I)~\cite{vitev}
and radiative plus collisional energy loss in the WHDG (II)~\cite{whdg2011} and CUJET1.0 (VII)~\cite{cujet} implementations describe reasonably well at the same time the charm and 
light-flavour suppression. 
While in the former calculation the medium density is tuned to describe the inclusive jet 
suppression at the LHC~\cite{vitevjet}, for the latter two it is extrapolated to LHC conditions starting from the 
value that describes the pion suppression at RHIC energy ($\sqrtsNN=200~\gev$). This could explain why these two models are somewhat low with respect to the charged-particle $\RAA$ data.
A model based on AdS/CFT drag coefficients (III)~\cite{horowitzAdSCFT} underestimates significantly the charm $\RAA$ and
has very limited predictive power for the light-flavour $\RAA$.

\section{Summary}
\label{sec:conclusions}
The first ALICE results on the nuclear modification factor $\RAA$ for charm
hadrons in Pb--Pb collisions at a centre-of-mass energy 
$\sqrtsNN=2.76~\tev$ indicate strong in-medium energy loss for charm quarks.
The $\rm D^0$, $\rm D^+$, and $\rm D^{*+}$ $\RAA$, 
measured for the first time  as a function of transverse momentum and 
centrality, is in the range 0.25--0.35 for \mbox{$5<\pt<16~\gev/c$} for the 
20\% most central collisions.
For $\pt$ below $5~\gev/c$, and towards peripheral collisions, 
there is a tendency for an increase of $\RAA$ for $\Dzero$ mesons. 

The suppression is almost 
as large as that observed for charged particles, which are mainly light-flavour
hadrons, with a possible indication, 
not fully significant with the present level of experimental uncertainties, 
of $\RAA^{\rm D}>\RAA^{\rm charged}$.
The expected effect of PDF nuclear shadowing is small ($<15\%$) above $\pt=6~\gev/c$, indicating
that the large measured suppression cannot be explained by initial-state effects.
Some of the pQCD models based on various implementations of parton energy loss 
succeed reasonably well at describing simultaneously the suppression of light 
flavour and charm hadrons.

The precision of the measurements will be improved in the future, using the large sample of Pb--Pb collisions recorded in 2011. In addition, p--Pb collision data expected in 2013 will provide insight on possible initial-state effects in 
the low-momentum region.

\vspace{1cm}


\newenvironment{acknowledgement}{\relax}{\relax}
\begin{acknowledgement}
\section*{Acknowledgements}
The ALICE collaboration would like to thank all its engineers and technicians for their invaluable contributions to the construction of the experiment and the CERN accelerator teams for the outstanding performance of the LHC complex.
The ALICE Collaboration would like to thank M.~Cacciari and H.~Spiesberger for 
providing the pQCD predictions used for the feed-down correction and the energy 
scaling, and the authors of the energy loss model calculations 
for making available their predictions for the nuclear modification factor.
\\
The ALICE collaboration acknowledges the following funding agencies for their support in building and
running the ALICE detector:
 \\
Calouste Gulbenkian Foundation from Lisbon and Swiss Fonds Kidagan, Armenia;
 \\
Conselho Nacional de Desenvolvimento Cient\'{\i}fico e Tecnol\'{o}gico (CNPq), Financiadora de Estudos e Projetos (FINEP),
Funda\c{c}\~{a}o de Amparo \`{a} Pesquisa do Estado de S\~{a}o Paulo (FAPESP);
 \\
National Natural Science Foundation of China (NSFC), the Chinese Ministry of Education (CMOE)
and the Ministry of Science and Technology of China (MSTC);
 \\
Ministry of Education and Youth of the Czech Republic;
 \\
Danish Natural Science Research Council, the Carlsberg Foundation and the Danish National Research Foundation;
 \\
The European Research Council under the European Community's Seventh Framework Programme;
 \\
Helsinki Institute of Physics and the Academy of Finland;
 \\
French CNRS-IN2P3, the `Region Pays de Loire', `Region Alsace', `Region Auvergne' and CEA, France;
 \\
German BMBF and the Helmholtz Association;
\\
General Secretariat for Research and Technology, Ministry of
Development, Greece;
\\
Hungarian OTKA and National Office for Research and Technology (NKTH);
 \\
Department of Atomic Energy and Department of Science and Technology of the Government of India;
 \\
Istituto Nazionale di Fisica Nucleare (INFN) of Italy;
 \\
MEXT Grant-in-Aid for Specially Promoted Research, Ja\-pan;
 \\
Joint Institute for Nuclear Research, Dubna;
 \\
National Research Foundation of Korea (NRF);
 \\
CONACYT, DGAPA, M\'{e}xico, ALFA-EC and the HELEN Program (High-Energy physics Latin-American--European Network);
 \\
Stichting voor Fundamenteel Onderzoek der Materie (FOM) and the Nederlandse Organisatie voor Wetenschappelijk Onderzoek (NWO), Netherlands;
 \\
Research Council of Norway (NFR);
 \\
Polish Ministry of Science and Higher Education;
 \\
National Authority for Scientific Research - NASR (Autoritatea Na\c{t}ional\u{a} pentru Cercetare \c{S}tiin\c{t}ific\u{a} - ANCS);
 \\
Federal Agency of Science of the Ministry of Education and Science of Russian Federation, International Science and
Technology Center, Russian Academy of Sciences, Russian Federal Agency of Atomic Energy, Russian Federal Agency for Science and Innovations and CERN-INTAS;
 \\
Ministry of Education of Slovakia;
 \\
Department of Science and Technology, South Africa;
 \\
CIEMAT, EELA, Ministerio de Educaci\'{o}n y Ciencia of Spain, Xunta de Galicia (Conseller\'{\i}a de Educaci\'{o}n),
CEA\-DEN, Cubaenerg\'{\i}a, Cuba, and IAEA (International Atomic Energy Agency);
 \\
Swedish Research Council (VR) and Knut $\&$ Alice Wallenberg
Foundation (KAW);
 \\
Ukraine Ministry of Education and Science;
 \\
United Kingdom Science and Technology Facilities Council (STFC);
 \\
The United States Department of Energy, the United States National
Science Foundation, the State of Texas, and the State of Ohio.
\end{acknowledgement}
\newpage
%
%
\appendix
\section{The ALICE Collaboration}
\label{app:collab}

\begingroup
\small
\begin{flushleft}
B.~Abelev\Irefn{org1234}\And
J.~Adam\Irefn{org1274}\And
D.~Adamov\'{a}\Irefn{org1283}\And
A.M.~Adare\Irefn{org1260}\And
M.M.~Aggarwal\Irefn{org1157}\And
G.~Aglieri~Rinella\Irefn{org1192}\And
A.G.~Agocs\Irefn{org1143}\And
A.~Agostinelli\Irefn{org1132}\And
S.~Aguilar~Salazar\Irefn{org1247}\And
Z.~Ahammed\Irefn{org1225}
% ...
\renewcommand\labelenumi{\textsuperscript{\theenumi}~}
\section*{Affiliation notes}
\renewcommand\theenumi{\roman{enumi}}
\begin{Authlist}
\item \Adef{M.V.Lomonosov Moscow State University, D.V.Skobeltsyn Institute of Nuclear Physics, Moscow, Russia}Also at: M.V.Lomonosov Moscow State University, D.V.Skobeltsyn Institute of Nuclear Physics, Moscow, Russia
\item \Adef{Institute of Nuclear Sciences, Belgrade, Serbia}Also at: "Vin\v{c}a" Institute of Nuclear Sciences, Belgrade, Serbia
\end{Authlist}
\section*{Collaboration Institutes}
\renewcommand\theenumi{\arabic{enumi}~}
\begin{Authlist}
\item \Idef{org1279}Benem\'{e}rita Universidad Aut\'{o}noma de Puebla, Puebla, Mexico
\item \Idef{org1220}Bogolyubov Institute for Theoretical Physics, Kiev, Ukraine
\item \Idef{org1262}Budker Institute for Nuclear Physics, Novosibirsk, Russia
\item \Idef{org1292}California Polytechnic State University, San Luis Obispo, California, United States
\item \Idef{org14939}Centre de Calcul de l'IN2P3, Villeurbanne, France
\item \Idef{org1197}Centro de Aplicaciones Tecnol\'{o}gicas y Desarrollo Nuclear (CEADEN), Havana, Cuba
\item \Idef{org1242}Centro de Investigaciones Energ\'{e}ticas Medioambientales y Tecnol\'{o}gicas (CIEMAT), Madrid, Spain
\item \Idef{org1244}Centro de Investigaci\'{o}n y de Estudios Avanzados (CINVESTAV), Mexico City and M\'{e}rida, Mexico
\item \Idef{org1335}Centro Fermi -- Centro Studi e Ricerche e Museo Storico della Fisica ``Enrico Fermi'', Rome, Italy
\item \Idef{org17347}Chicago State University, Chicago, United States
\item \Idef{org1288}Commissariat \`{a} l'Energie Atomique, IRFU, Saclay, France
\item \Idef{org1294}Departamento de F\'{\i}sica de Part\'{\i}culas and IGFAE, Universidad de Santiago de Compostela, Santiago de Compostela, Spain
\item \Idef{org1106}Department of Physics Aligarh Muslim University, Aligarh, India
\item \Idef{org1121}Department of Physics and Technology, University of Bergen, Bergen, Norway
\item \Idef{org1162}Department of Physics, Ohio State University, Columbus, Ohio, United States
\item \Idef{org1300}Department of Physics, Sejong University, Seoul, South Korea
\item \Idef{org1268}Department of Physics, University of Oslo, Oslo, Norway
\item \Idef{org1145}Dipartimento di Fisica dell'Universit\`{a} and Sezione INFN, Cagliari, Italy
\item \Idef{org1270}Dipartimento di Fisica dell'Universit\`{a} and Sezione INFN, Padova, Italy
\item \Idef{org1315}Dipartimento di Fisica dell'Universit\`{a} and Sezione INFN, Trieste, Italy
\item \Idef{org1132}Dipartimento di Fisica dell'Universit\`{a} and Sezione INFN, Bologna, Italy
\item \Idef{org1285}Dipartimento di Fisica dell'Universit\`{a} `La Sapienza' and Sezione INFN, Rome, Italy
\item \Idef{org1154}Dipartimento di Fisica e Astronomia dell'Universit\`{a} and Sezione INFN, Catania, Italy
\item \Idef{org1290}Dipartimento di Fisica `E.R.~Caianiello' dell'Universit\`{a} and Gruppo Collegato INFN, Salerno, Italy
\item \Idef{org1312}Dipartimento di Fisica Sperimentale dell'Universit\`{a} and Sezione INFN, Turin, Italy
\item \Idef{org1103}Dipartimento di Scienze e Tecnologie Avanzate dell'Universit\`{a} del Piemonte Orientale and Gruppo Collegato INFN, Alessandria, Italy
\item \Idef{org1114}Dipartimento Interateneo di Fisica `M.~Merlin' and Sezione INFN, Bari, Italy
\item \Idef{org1237}Division of Experimental High Energy Physics, University of Lund, Lund, Sweden
\item \Idef{org1192}European Organization for Nuclear Research (CERN), Geneva, Switzerland
\item \Idef{org1227}Fachhochschule K\"{o}ln, K\"{o}ln, Germany
\item \Idef{org1122}Faculty of Engineering, Bergen University College, Bergen, Norway
\item \Idef{org1136}Faculty of Mathematics, Physics and Informatics, Comenius University, Bratislava, Slovakia
\item \Idef{org1274}Faculty of Nuclear Sciences and Physical Engineering, Czech Technical University in Prague, Prague, Czech Republic
\item \Idef{org1229}Faculty of Science, P.J.~\v{S}af\'{a}rik University, Ko\v{s}ice, Slovakia
\item \Idef{org1184}Frankfurt Institute for Advanced Studies, Johann Wolfgang Goethe-Universit\"{a}t Frankfurt, Frankfurt, Germany
\item \Idef{org1215}Gangneung-Wonju National University, Gangneung, South Korea
\item \Idef{org1212}Helsinki Institute of Physics (HIP) and University of Jyv\"{a}skyl\"{a}, Jyv\"{a}skyl\"{a}, Finland
\item \Idef{org1203}Hiroshima University, Hiroshima, Japan
\item \Idef{org1329}Hua-Zhong Normal University, Wuhan, China
\item \Idef{org1254}Indian Institute of Technology, Mumbai, India
\item \Idef{org36378}Indian Institute of Technology Indore (IIT), Indore, India
\item \Idef{org1266}Institut de Physique Nucl\'{e}aire d'Orsay (IPNO), Universit\'{e} Paris-Sud, CNRS-IN2P3, Orsay, France
\item \Idef{org1277}Institute for High Energy Physics, Protvino, Russia
\item \Idef{org1249}Institute for Nuclear Research, Academy of Sciences, Moscow, Russia
\item \Idef{org1320}Nikhef, National Institute for Subatomic Physics and Institute for Subatomic Physics of Utrecht University, Utrecht, Netherlands
\item \Idef{org1250}Institute for Theoretical and Experimental Physics, Moscow, Russia
\item \Idef{org1230}Institute of Experimental Physics, Slovak Academy of Sciences, Ko\v{s}ice, Slovakia
\item \Idef{org1127}Institute of Physics, Bhubaneswar, India
\item \Idef{org1275}Institute of Physics, Academy of Sciences of the Czech Republic, Prague, Czech Republic
\item \Idef{org1139}Institute of Space Sciences (ISS), Bucharest, Romania
\item \Idef{org27399}Institut f\"{u}r Informatik, Johann Wolfgang Goethe-Universit\"{a}t Frankfurt, Frankfurt, Germany
\item \Idef{org1185}Institut f\"{u}r Kernphysik, Johann Wolfgang Goethe-Universit\"{a}t Frankfurt, Frankfurt, Germany
\item \Idef{org1177}Institut f\"{u}r Kernphysik, Technische Universit\"{a}t Darmstadt, Darmstadt, Germany
\item \Idef{org1256}Institut f\"{u}r Kernphysik, Westf\"{a}lische Wilhelms-Universit\"{a}t M\"{u}nster, M\"{u}nster, Germany
\item \Idef{org1246}Instituto de Ciencias Nucleares, Universidad Nacional Aut\'{o}noma de M\'{e}xico, Mexico City, Mexico
\item \Idef{org1247}Instituto de F\'{\i}sica, Universidad Nacional Aut\'{o}noma de M\'{e}xico, Mexico City, Mexico
\item \Idef{org23333}Institut of Theoretical Physics, University of Wroclaw
\item \Idef{org1308}Institut Pluridisciplinaire Hubert Curien (IPHC), Universit\'{e} de Strasbourg, CNRS-IN2P3, Strasbourg, France
\item \Idef{org1182}Joint Institute for Nuclear Research (JINR), Dubna, Russia
\item \Idef{org1143}KFKI Research Institute for Particle and Nuclear Physics, Hungarian Academy of Sciences, Budapest, Hungary
\item \Idef{org1199}Kirchhoff-Institut f\"{u}r Physik, Ruprecht-Karls-Universit\"{a}t Heidelberg, Heidelberg, Germany
\item \Idef{org20954}Korea Institute of Science and Technology Information, Daejeon, South Korea
\item \Idef{org1160}Laboratoire de Physique Corpusculaire (LPC), Clermont Universit\'{e}, Universit\'{e} Blaise Pascal, CNRS--IN2P3, Clermont-Ferrand, France
\item \Idef{org1194}Laboratoire de Physique Subatomique et de Cosmologie (LPSC), Universit\'{e} Joseph Fourier, CNRS-IN2P3, Institut Polytechnique de Grenoble, Grenoble, France
\item \Idef{org1187}Laboratori Nazionali di Frascati, INFN, Frascati, Italy
\item \Idef{org1232}Laboratori Nazionali di Legnaro, INFN, Legnaro, Italy
\item \Idef{org1125}Lawrence Berkeley National Laboratory, Berkeley, California, United States
\item \Idef{org1234}Lawrence Livermore National Laboratory, Livermore, California, United States
\item \Idef{org1251}Moscow Engineering Physics Institute, Moscow, Russia
\item \Idef{org1140}National Institute for Physics and Nuclear Engineering, Bucharest, Romania
\item \Idef{org1165}Niels Bohr Institute, University of Copenhagen, Copenhagen, Denmark
\item \Idef{org1109}Nikhef, National Institute for Subatomic Physics, Amsterdam, Netherlands
\item \Idef{org1283}Nuclear Physics Institute, Academy of Sciences of the Czech Republic, \v{R}e\v{z} u Prahy, Czech Republic
\item \Idef{org1264}Oak Ridge National Laboratory, Oak Ridge, Tennessee, United States
\item \Idef{org1189}Petersburg Nuclear Physics Institute, Gatchina, Russia
\item \Idef{org1170}Physics Department, Creighton University, Omaha, Nebraska, United States
\item \Idef{org1157}Physics Department, Panjab University, Chandigarh, India
\item \Idef{org1112}Physics Department, University of Athens, Athens, Greece
\item \Idef{org1152}Physics Department, University of Cape Town, iThemba LABS, Cape Town, South Africa
\item \Idef{org1209}Physics Department, University of Jammu, Jammu, India
\item \Idef{org1207}Physics Department, University of Rajasthan, Jaipur, India
\item \Idef{org1200}Physikalisches Institut, Ruprecht-Karls-Universit\"{a}t Heidelberg, Heidelberg, Germany
\item \Idef{org1325}Purdue University, West Lafayette, Indiana, United States
\item \Idef{org1281}Pusan National University, Pusan, South Korea
\item \Idef{org1176}Research Division and ExtreMe Matter Institute EMMI, GSI Helmholtzzentrum f\"ur Schwerionenforschung, Darmstadt, Germany
\item \Idef{org1334}Rudjer Bo\v{s}kovi\'{c} Institute, Zagreb, Croatia
\item \Idef{org1298}Russian Federal Nuclear Center (VNIIEF), Sarov, Russia
\item \Idef{org1252}Russian Research Centre Kurchatov Institute, Moscow, Russia
\item \Idef{org1224}Saha Institute of Nuclear Physics, Kolkata, India
\item \Idef{org1130}School of Physics and Astronomy, University of Birmingham, Birmingham, United Kingdom
\item \Idef{org1338}Secci\'{o}n F\'{\i}sica, Departamento de Ciencias, Pontificia Universidad Cat\'{o}lica del Per\'{u}, Lima, Peru
\item \Idef{org1316}Sezione INFN, Trieste, Italy
\item \Idef{org1271}Sezione INFN, Padova, Italy
\item \Idef{org1313}Sezione INFN, Turin, Italy
\item \Idef{org1286}Sezione INFN, Rome, Italy
\item \Idef{org1146}Sezione INFN, Cagliari, Italy
\item \Idef{org1133}Sezione INFN, Bologna, Italy
\item \Idef{org1115}Sezione INFN, Bari, Italy
\item \Idef{org1155}Sezione INFN, Catania, Italy
\item \Idef{org1322}Soltan Institute for Nuclear Studies, Warsaw, Poland
\item \Idef{org36377}Nuclear Physics Group, STFC Daresbury Laboratory, Daresbury, United Kingdom
\item \Idef{org1258}SUBATECH, Ecole des Mines de Nantes, Universit\'{e} de Nantes, CNRS-IN2P3, Nantes, France
\item \Idef{org1304}Technical University of Split FESB, Split, Croatia
\item \Idef{org1168}The Henryk Niewodniczanski Institute of Nuclear Physics, Polish Academy of Sciences, Cracow, Poland
\item \Idef{org17361}The University of Texas at Austin, Physics Department, Austin, TX, United States
\item \Idef{org1173}Universidad Aut\'{o}noma de Sinaloa, Culiac\'{a}n, Mexico
\item \Idef{org1296}Universidade de S\~{a}o Paulo (USP), S\~{a}o Paulo, Brazil
\item \Idef{org1149}Universidade Estadual de Campinas (UNICAMP), Campinas, Brazil
\item \Idef{org1239}Universit\'{e} de Lyon, Universit\'{e} Lyon 1, CNRS/IN2P3, IPN-Lyon, Villeurbanne, France
\item \Idef{org1205}University of Houston, Houston, Texas, United States
\item \Idef{org20371}University of Technology and Austrian Academy of Sciences, Vienna, Austria
\item \Idef{org1222}University of Tennessee, Knoxville, Tennessee, United States
\item \Idef{org1310}University of Tokyo, Tokyo, Japan
\item \Idef{org1318}University of Tsukuba, Tsukuba, Japan
\item \Idef{org21360}Eberhard Karls Universit\"{a}t T\"{u}bingen, T\"{u}bingen, Germany
\item \Idef{org1225}Variable Energy Cyclotron Centre, Kolkata, India
\item \Idef{org1306}V.~Fock Institute for Physics, St. Petersburg State University, St. Petersburg, Russia
\item \Idef{org1323}Warsaw University of Technology, Warsaw, Poland
\item \Idef{org1179}Wayne State University, Detroit, Michigan, United States
\item \Idef{org1260}Yale University, New Haven, Connecticut, United States
\item \Idef{org1332}Yerevan Physics Institute, Yerevan, Armenia
\item \Idef{org15649}Yildiz Technical University, Istanbul, Turkey
\item \Idef{org1301}Yonsei University, Seoul, South Korea
\item \Idef{org1327}Zentrum f\"{u}r Technologietransfer und Telekommunikation (ZTT), Fachhochschule Worms, Worms, Germany
\end{Authlist}
\endgroup


\end{document}